\documentclass[preprint,12pt]{elsarticle}
\usepackage{lineno,hyperref}
\modulolinenumbers[1]
\graphicspath{{figures/}}
\usepackage{fullpage}

\usepackage{bm}

\journal{Nuclear Instruments and Methods in Physics Research A}

\newcommand{\moller}{M{\o}ller}
\newcommand{\prex}{PREX-2}
\newcommand{\crex}{CREX}

\begin{document}

\begin{frontmatter}

\title{Precision \moller\ Polarimetry for \prex\ and \crex}
%\tnotetext[mytitlenote]{Fully documented templates are available in the elsarticle package on \href{http://www.ctan.org/tex-archive/macros/latex/contrib/elsarticle}{CTAN}.}

%% Group authors per affiliation:
%\author{Elsevier\fnref{myfootnote}}
%\address{Radarweg 29, Amsterdam}
%\fntext[myfootnote]{Since 1880.}

%% or include affiliations in footnotes:
\author[temple,syracuse]{D. E. King}
\author[jlab,temple]{D. C. Jones}
\author[sunysb,msstate]{C. Gal}
\author[jlab]{D. Gaskell}
\author[jlab]{W. Henry}
\author[temple]{A.\ D.\ Kaplan}
\author[temple]{J. Napolitano}
\author[sunysb,msstate]{S. Park}
\author[virginia]{K.D. Paschke}
\author[kharkov]{R. Pomatsalyuk}
\author[syracuse]{P. A. Souder}

%%%%% CORRESPONDING AUTHOR EXAMPLE SETUP -- DON'T DELETE YET
%\author[mysecondaryaddress]{Global Customer Service\corref{mycorrespondingauthor}}
%\cortext[mycorrespondingauthor]{Corresponding author}
%\ead{support@elsevier.com}

\address[temple]{Temple University, Philadelphia, PA, 19122}
\address[syracuse]{Syracuse University, Syracuse, NY, 13244, USA}
\address[jlab]{Thomas Jefferson National Accelerator Facility, Newport News, VA, 23606}
\address[sunysb]{State University of New York, Stony Brook, NY, 11794}
\address[msstate]{Mississippi State University, MS, 39762}
\address[virginia]{University of Virginia, Charlottesville, VA 22904 USA}
\address[kharkov]{Kharkov Institute of Physics and Technology, Kharkov 61108, Ukraine}

\begin{abstract}
The \prex\ and \crex\ experiments in Hall~A at Jefferson Lab are precision measurements of parity violating elastic electron scattering from complex nuclei. One requirement was that the incident electron beam polarization, typically $\approx90\%$, be known with 1\% precision. We commissioned and operated a \moller\ polarimeter on the beam line that exceeds this requirement, achieving a precision of 0.89\% for \prex, and 0.85\% for \crex. The uncertainty is purely systematic, accumulated from several different sources, but dominated by our knowledge of the target polarization. Our analysis also demonstrates the need for accurate atomic wave functions in order to correct for the Levchuk Effect. We describe the details of the polarimeter operation and analysis, as well as (for \crex) a comparison to results from a different polarimeter based on Compton scattering.
\end{abstract}

\begin{keyword}
\texttt{elsarticle.cls}\sep \LaTeX\sep Elsevier \sep template
\MSC[2010] 00-01\sep  99-00
\end{keyword}

\end{frontmatter}

%\linenumbers

%\centerline{This file processed on \today}

%[Internal] Link to Don's proposed outline \href{https://docs.google.com/document/d/1UFvS-mOb1G9W-fTp5n868AwIRmsRSHQYA05CoeXqpM0/edit?usp=sharing}{here}.

\section{Introduction}
M{\o}ller polarimetry has proven to be a very useful technique for measuring the polarization of GeV electron beams for nuclear and high energy physics experiments \cite{Magee:2016xqx,Glamazdin:1999gg,Steiner:1998gf,Band:1997ee,Arrington:1992xu}.
%The approach makes use of the spin asymmetry $A_{zz}(\theta)$ in the scattering of longitudinally polarized beam electrons from longitudinally polarized target electrons, which can be calculated to high precision in Quantum Electrodynamics (QED) as a function of center-of-mass scattering angle $\theta$. 
The approach takes advantage of the spin-correlated asymmetry of \moller\ scattering which can be calculated to high precision in Quantum Electrodynamics (QED) as a function of the center-of-mass scattering angle $\theta$ and is an ideal way to determine the polarization of a longitudinally polarized electron beam polarization utilizing a longitudinally polarized target. 
The beam polarization is inferred from the target polarization and the average analyzing power of the spectrometer used to momentum-analyze the scattered electron(s). That is
\begin{equation}
A_{\rm meas}=\frac{R_{\uparrow\uparrow}-R_{\downarrow\uparrow}}{R_{\uparrow\uparrow}+R_{\downarrow\uparrow}}=-P_{\rm beam}P_{\rm target}\langle A_{zz}\rangle,
%A_{\rm meas}=\frac{R_{+}-R_{-}}{R_{+}+R_{-}}=-P_{\rm beam}P_{\rm target}\langle A_{zz}\rangle,
\label{eq:MeasuredAsymmetry}
\end{equation}
where $R_{\uparrow\uparrow}$ and $R_{\downarrow\uparrow}$ are the measured rates for parallel and anti-parallel beam-target spin states, $P_{\rm target}$ is the electron spin polarization of the target, $\langle A_{zz}\rangle$ is the fundamental longitudinal analyzing power of the \moller\ scattering reaction averaged over the instrumental acceptance, and $P_{\rm beam}$ is the electron beam polarization.
%where $R_\pm$ are the measured rates for positive and negative helicity states of the beam, $P_{\rm target}$ is the electron spin polarization of the target, $\langle A_{zz}\rangle$ is the fundamental longitudinal analyzing power of the \moller\ scattering reaction averaged over the instrumental acceptance, and $P_{\rm beam}$ is the electron beam polarization.

The fundamental spin asymmetry, to lowest order in QED, is
\begin{equation}
A_{zz}(\theta)=\frac{(7+\cos^2\theta)\sin^2\theta}{(3+\cos^2\theta)^2}.
\label{eq:TheoreticalAsymmetry}
\end{equation}
The large value of $A_{zz}^{\rm max}=7/9$ at $\theta=90^\circ$ is an important reason that this is a useful technique. A large analyzing power minimizes the length of time needed to acquire sufficient statistical precision for $A_{\rm meas}$ in Equation~\ref{eq:MeasuredAsymmetry}. Therefore, \moller\ polarimeter spectrometers are typically designed to accept some range of $\theta$ around $90^\circ$.

It is generally not a challenge to acquire data at a high enough rate so that the necessary statistical precision can be achieved in a reasonable amount of time. Consequently, the precision with which one can determine $P_{\rm beam}$ using Equation~\ref{eq:MeasuredAsymmetry} is typically constrained by systematic uncertainties.

In fact, there are a wide variety of systematic uncertainty contributions to each of $A_{\rm meas}$, $P_{\rm target}$, and $\langle A_{zz}\rangle$. For $A_{\rm meas}$, these include the effects of helicity-correlated beam asymmetries, dead time corrections, and potential backgrounds from other sources of polarized beam (see Section \ref{sec:leakage}). The averaging process required to determine $\langle A_{zz}\rangle$ depends on the accuracy of the simulation programs, and the effect of the motion of electrons in the polarized electron target, a phenomenon known as the Levchuk Effect~\cite{Levchuk1994}.

Systematic uncertainties to $P_{\rm target}$ are especially important. All operational \moller\ polarimeters to date make use of a magnetized ferromagnetic foil as the target. Earlier designs used a high permeability alloy, held at some angle to the incident electron beam, polarized in a holding magnetic field of hundreds of Gauss. Although such magnetic fields are not difficult to achieve with resistive coils, the systematic uncertainties on $P_{\rm target}$ were relatively large. This was because of both the transverse components in the tilted foil, necessary since the magnetization is in the plane of the foil, and the difficulty in extracting the actual spin polarization from the magnetization in these complex alloys.

The new generation of \moller\ polarimeters, based on a technique developed for Hall~C at Jefferson Lab~\cite{bever:1997kvc}, instead use a pure iron foil target, polarized perpendicular to its plane using a high magnetic field to saturate the iron. Magnetic fields in excess of 2~T are needed in this technique, so one resorts to holding fields supplied by superconducting magnets. The germane magnetic properties of a pure iron target allow the polarization $P_{\rm target}$ to be determined to 0.24\% precision.\cite{Jones:2022jhy} This is the ultimate limiting precision for \moller\ polarimeters that make use of this kind of polarized target.

This paper describes the operation and results with the \moller\ polarimeter in Hall~A at Jefferson Lab during the \prex~\cite{PREX2_2021} and \crex~\cite{CREX_2022} experiments using the Continuous Electron Beam Accelerator Facility (CEBAF). These experiments used parity violating electron scattering to determine the neutron ``skin thickness'' in the isotopes $^{208}$Pb and $^{48}$Ca, respectively. The contribution of the neutron skin to the asymmetry is relatively small, necessitating a high precision measurement to extract the skin thickness with moderate precision.

The rest of this paper is organized as follows. Section~\ref{sec:Spectrometer} describes the Hall~A \moller\ polarimeter and gives the results of our measurements of the polarization, including a description of the techniques used in \prex\ and \crex, and the associated systematic uncertainties. Section~\ref{sec:measurement} gives details on ``measurement'' uncertainties, associated with how well we know various quantities at the time of the measurement. Section~\ref{sec:extrapolation} then discusses ``extrapolated'' uncertainties which arise from the fact that the conditions under which the polarization measurements were made are different from the operating conditions of the main experiment necessitating an extrapolation from the measurements to the conditions and time at which the data were taken. Finally, we offer some conclusions and prospects for future measurements.
\begin{figure*}[t!]
    \centering
    \includegraphics[width=\textwidth]{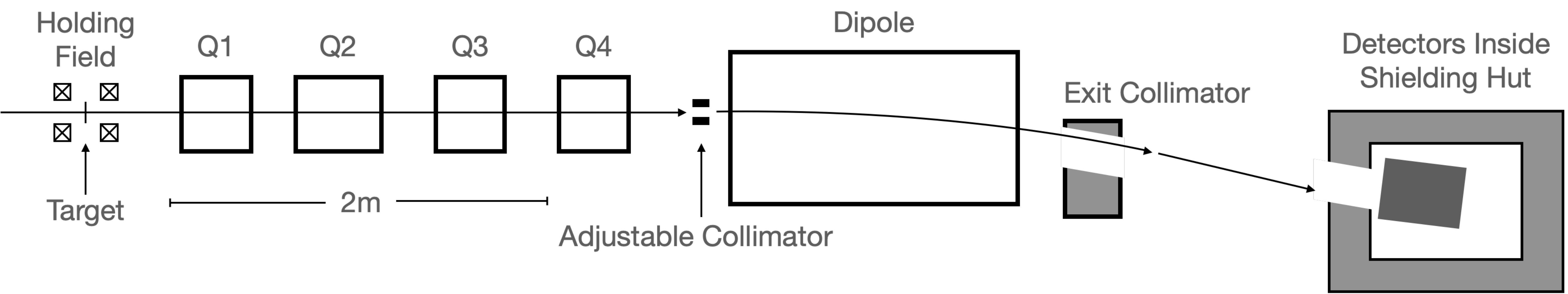}
    \caption{Elevation view schematic of the \moller\ polarimeter used for the \prex\ and \crex\ experiments in Hall~A at Jefferson Lab. The spectrometer consists of four focussing quadrupole magnets labeled Q1, Q2, Q3, and Q4, and a single dipole magnet for momentum analysis. For the measurements described in this paper, run at relatively low energies, Q3 was turned off, but the four-quadrupole configuration is critical for the higher energies available from the Continuous Electron Beam Accelerator Facility (CEBAF).
    \label{fig:Schematic}}
\end{figure*}

\section{Spectrometer and Measurements}
\label{sec:Spectrometer}

Figure~\ref{fig:Schematic} shows a schematic of the \moller\ polarimeter as configured for \prex\ and \crex, drawn to the indicated scale. Polarized target electrons are provided by a pure iron target foil, with its plane perpendicular to the beam axis, magnetized to saturation by a superconducting Helmholtz coil up to 4~T with field along the beam axis. After emerging from the target, a series of four quadrupole magnets align the two outgoing electrons with center-of-mass angle near $90^\circ$ to be approximately parallel before entering the dipole magnet. An adjustable collimator at the entrance to the dipole magnet is used to restrict the azimuthal angle of the scattering with respect to the beam axis. The electrons are collimated at the dipole exit before drifting $\approx1$~m and entering a shielded detector hut containing the electron detectors. An iron tube provides magnetic shielding for the primary beam as it passes through the dipole on its way to the primary target pivot and beam dump.

The electron detectors consist of four lead blocks, each $9\times15$~cm$^2$ and 30~cm long, infused with scintillating fibers. The fibers extend out the back of the detectors and, for each of the four blocks, are gathered into two bundles, one each on the top and bottom. Each bundle is connected to one photo-multiplier tube (PMT), effectively splitting the block into two detector blocks. The sum of the four left and four right PMTs are multiplexed separately and discriminated to form a simple coincidence trigger. Each PMT is also fed into an analog-to-digital converter (ADC) for gain matching and setting the discriminator threshold. This setup effectively eliminates backgrounds from other electron scattering processes, mainly Mott scattering from the iron nuclei.

We developed a complete simulation of the target, spectrometer, and detectors using {\sc Geant4}.~\cite{GEANT4:2002zbu,Allison:2006ve,Allison:2016lfl} The simulation was first validated against an earlier simulation~\cite{Glamazdin:1999gg} of the spectrometer, and then with extensive comparisons to data, some of which are described below, mainly in Section~\ref{subsec:CREX} which discusses data taking during \crex. For more details, see~\cite{EricKingThesis}.

For \prex\ and \crex, which ran at beam energies of 0.95~GeV and 2.2~GeV, respectively, quadrupole magnet Q3 was turned off. For operation at high energies, up to 11~GeV, the first two quadrupoles are used to defocus the electron pair so that they enter Q3 farther from the beam axis and can therefore be more effectively focused into the dipole. Indeed, the configuration in Figure~\ref{fig:Schematic} is an evolution from the three-quadrupole system~\cite{Glamazdin:1999gg} used for beam energies below 6~GeV.

The \prex\ experiment ran in the summer of 2019, and \crex\ ran in the winter, summer, and fall of 2020, with a long break in the spring.
% due to the COVID-19 pandemic.
We took \moller\ polarimeter measurements approximately every week during \prex\ and every 2-3 weeks during \crex. The polarimeter is located upstream of the main target in Hall A, and the polarimeter quadrupoles are incorporated into the beam line tune. Therefore, each polarimetry measurement required us to restore the magnetic tune of the spectrometer, which involved matching the coincident electron rates to the simulation as a function of the field setting in one or another of the spectrometer magnets.  Also, given the different beam energies, there needed to be significant changes to the magnetic tune of the spectrometer between \prex\ and \crex.

We note that \moller\ polarimetry with iron foil targets is beam destructive, so it can only be deployed interspersed between data runs for the main experiment. However, a different, non-beam destructive polarimeter based on Compton scattering was deployed during \crex.~\cite{AllisonZecThesis} This allowed cross checks between the beam polarization measurements for the two techniques.

\subsection{Determination of the Analyzing Power}

An accurate and precise determination of the average analyzing power is critical to a precise measurement of the polarization. We achieved this using a simulation that was confirmed by taking data in different configurations. This includes verification of the Levchuk Effect~\cite{Levchuk1994}, with which we find excellent agreement after using improved atomic wave functions.

\begin{figure*}
\begin{center}
\includegraphics[width=0.49\textwidth]{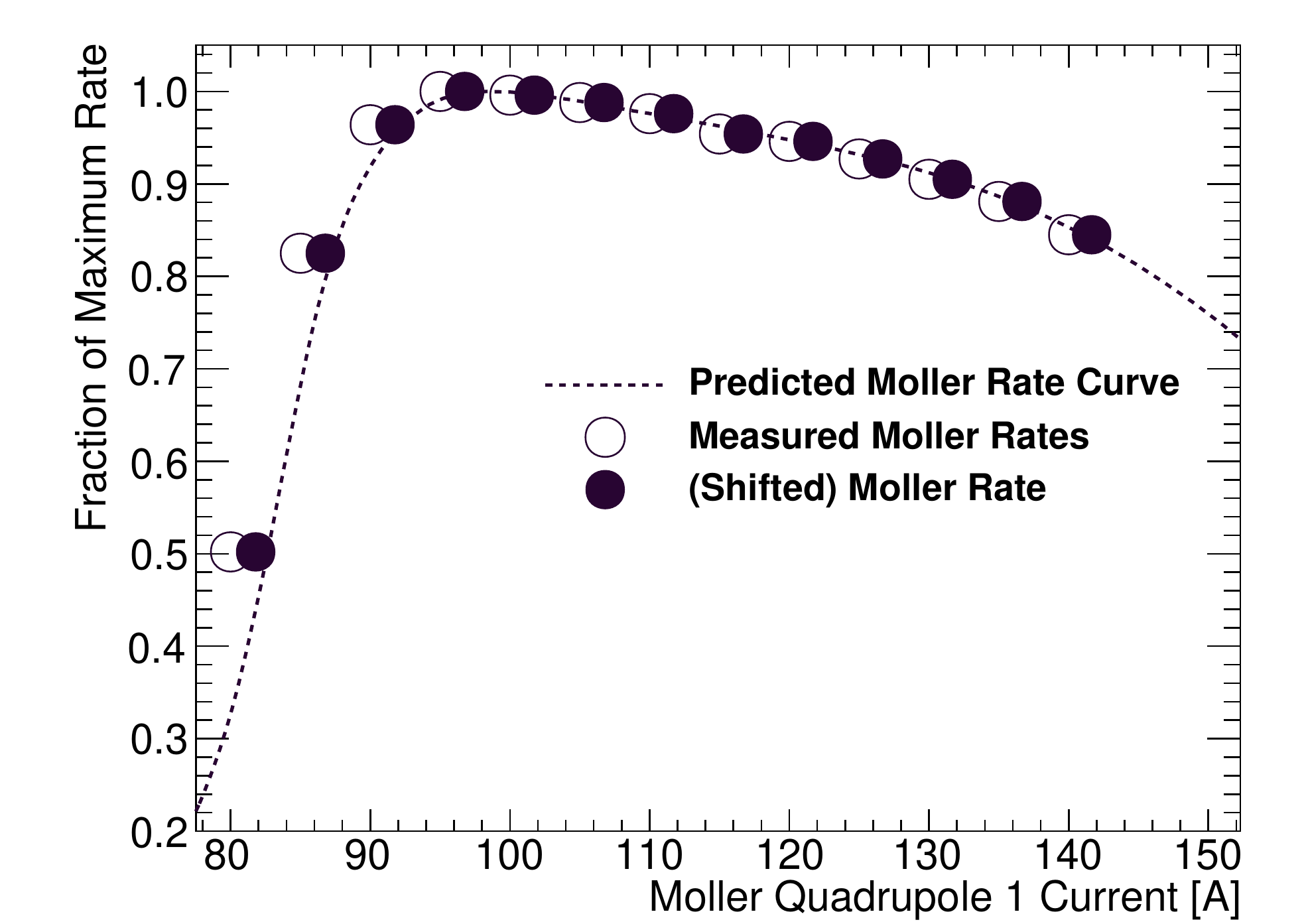}
\includegraphics[width=0.49\textwidth]{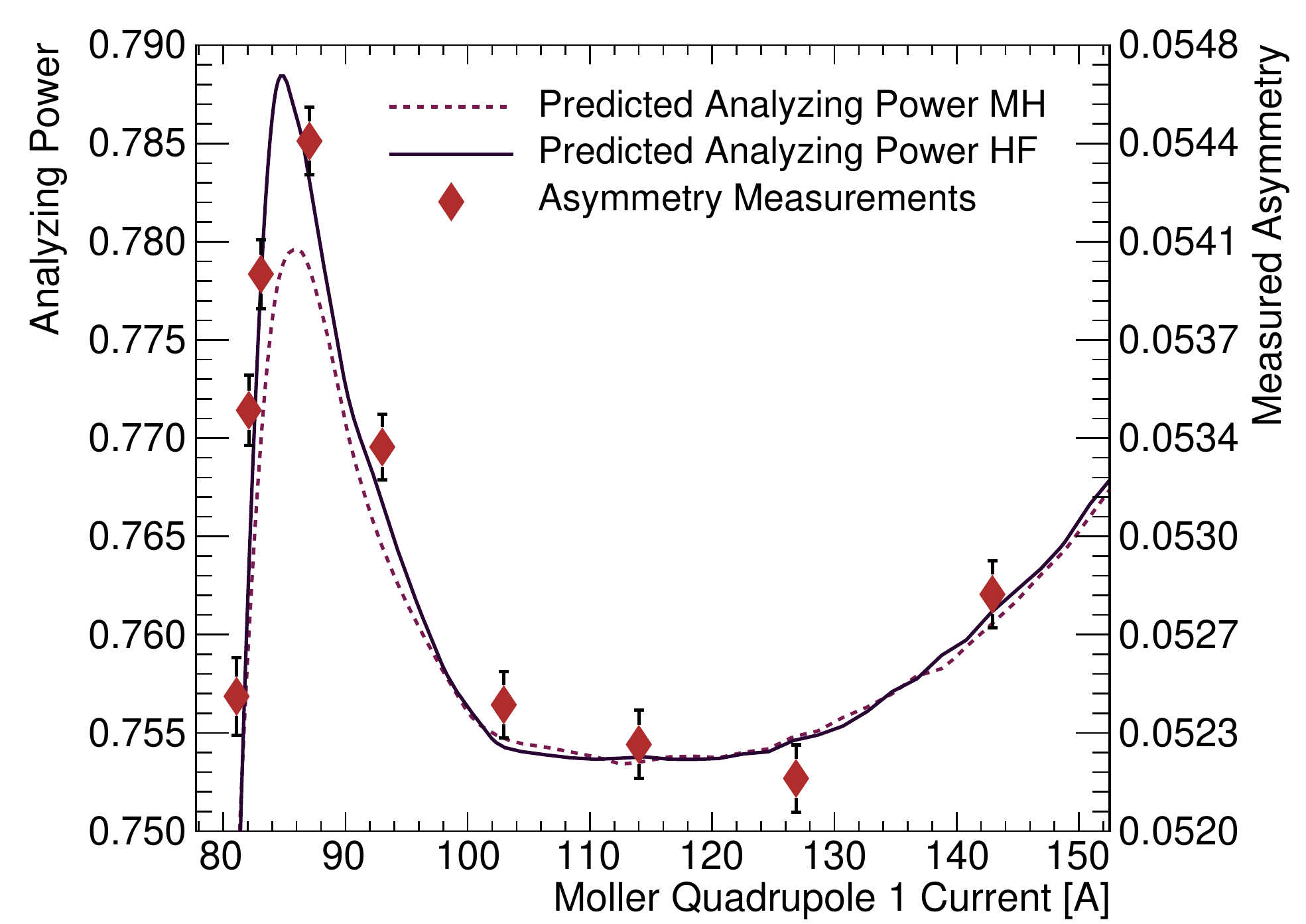}
\end{center}
\caption{Coincidence scattering rate (left) and analyzing power (right), as a function of magnet current for Q1 (Figure~\ref{fig:Schematic}) for \moller\ polarimetry during \crex. The predictions are from our {\sc Geant4} simulation. The rate curve shows that we needed to shift the magnetic field slightly relative to the nominal field given by the power supply setting. The analyzing power curve is normalized vertically to agree with measurements, thereby determining the actual beam polarization, and the two simulations refer to atomic wave functions from hydrogen modified for the iron nucleus (MH) and for a more realistic Hartree Fock (HF) calculation. The peak in analyzing power near 90~A is due to the Levchuk Effect. The error bars on the analyzing power are purely statistical.
\label{fig:ScansCREX}}
\end{figure*}

Figure~\ref{fig:ScansCREX}
shows the procedures we used to determine $\langle A_{zz}\rangle$. First we measured the coincidence rate as a function of magnet settings. The figure shows this as a function of quadrupole Q1 field strength (see Figure~\ref{fig:Schematic}). We match these rates to what we predict using the {\sc Geant4} Monte Carlo simulation. This led us to make a $\approx2\%$ correction to the nominal magnetic field indicated by the power supply current, after consideration of which we get excellent agreement with the simulation. We note that this 2\% correction is consistent with accepted uncertainties in the magnet calibration.

We then use the simulation to predict the (effective) analyzing power $\langle A_{zz}\rangle$, again as a function of magnet settings, including the current-to-field adjustment indicated from the rate measurements. The simulation includes the momentum distributions of the atomic electrons, as well as averaging Eq.~\ref{eq:TheoreticalAsymmetry} (plus radiative corrections~\cite{Afanasev:2006xs}) over the angular acceptance of the spectrometer. We then measure the \moller\ asymmetry in Eq.~\ref{eq:MeasuredAsymmetry} and plot it as $\langle A_{zz}\rangle$ by including our knowledge of the target polarization and a fitted value for the beam polarization.

The large enhancement to the analyzing power in the region of 90~A is due to the Levchuk Effect~\cite{Levchuk1994}. Only the outer, less tightly-bound electrons in the iron atom are polarized. However, \moller\ scattering obviously occurs from all the atomic electrons. The inner electrons have a momentum distribution with a long tail to higher momenta, giving a wider angular distribution to the scattered electron pair. Our Monte Carlo simulation work shows that the region of enhanced analyzing power is a direct result of the decreased acceptance of \moller\ scatters from inner-bound electrons whose trajectories were impacted by the Levchuk Effect. Thus this produces an increased acceptance bias towards scatters from the unpolarized outer electrons.

This enhancement was predicted with $\approx30\%$ accuracy with a simulation that included iron electron momentum distributions that came from simple hydrogen atom wave functions modified for $Z=26$~\cite{Swartz1995}. The agreement became much closer after we used wave functions from a proper Hartree Fock calculation. (Details are in Section~\ref{subsec:Kaplan} below.) This gives us good confidence that we can make the Levchuk Effect correction with a precision better than 10\%. We nevertheless take production polarimetry data in the ``flat'' region near 115~A where the correction is small.

\subsection{\moller\ polarimetry during \prex}
\label{subsec:PREX}
The \prex\ experiment took data from July to early September of 2019. During July and early August we commissioned the \moller\  polarimeter studying sensitivity to detector thresholds, spatial distributions on the detectors, energy spectra, and detector high voltage settings. Prior to taking production data for polarization measurements, we measured the coincidence scattering rate and asymmetry as a function of magnet settings, and compared these to simulation. This allowed us to finalize settings for production data which were insensitive to precise knowledge of the magnetic fields while maximizing count rate. See Fig.~\ref{fig:ScansCREX} and associated text.

\begin{figure*}
\centerline{\includegraphics[width=0.91\textwidth]{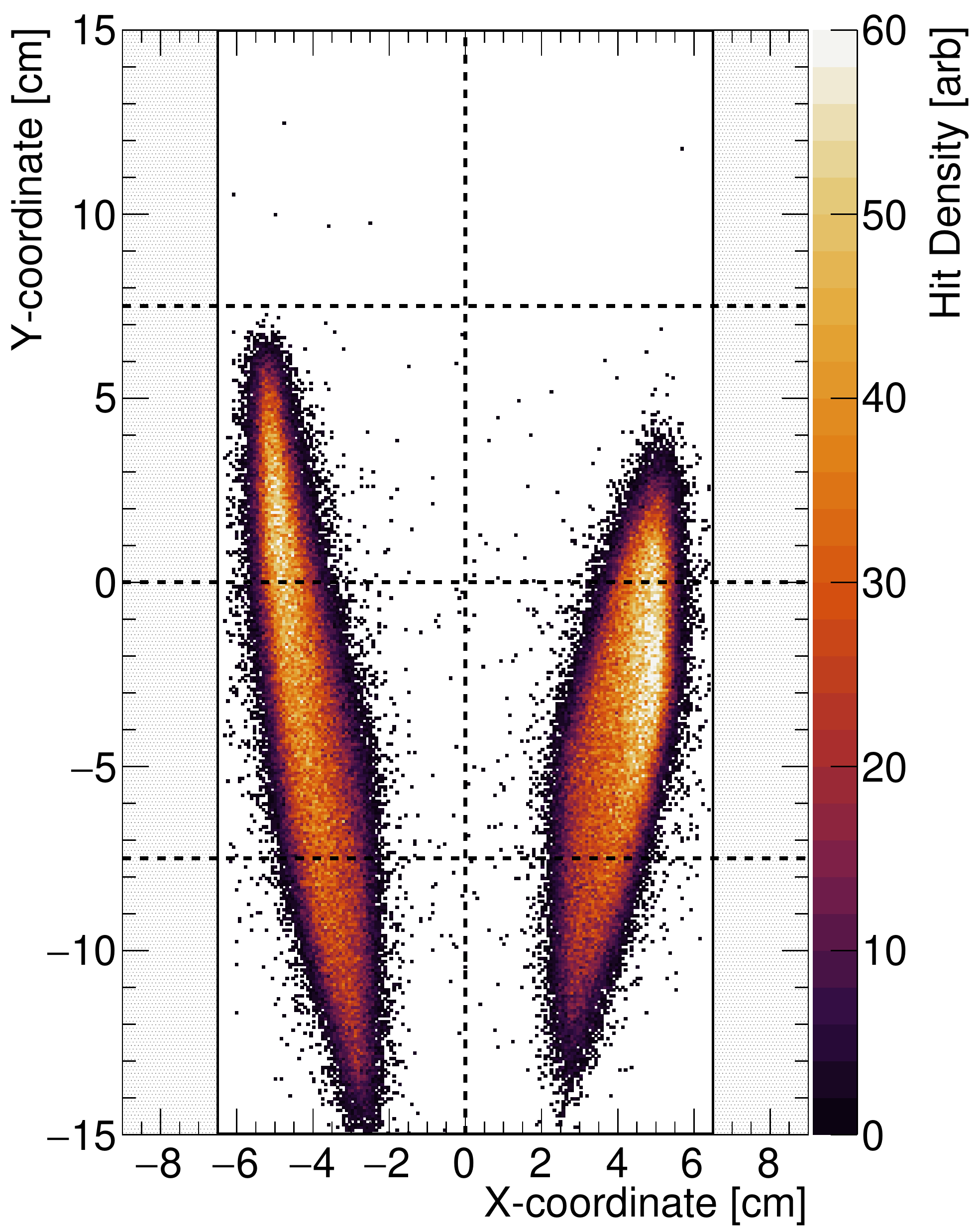}}
\caption{Positions of coincident \moller\ scattered electrons on the detector face, from our {\sc Geant4} simulation for parameters corresponding to \prex\ running conditions. The extent of the axes correspond to the physical size of the calorimeter. The dashed lines delineate the eight separate detector blocks. The solid vertical lines at $x=\pm 6.5$ cm correspond to the collimator at the entrance of the detector hut.  The left-right asymmetry is due to bending in the Helmholtz coil target holding field.
\label{fig:StripePREX}}
\end{figure*}

Figure~\ref{fig:StripePREX} shows a heat map of where the \moller\ coincidence events impinge on the detector face, based on our {\sc Geant4} simulation under \prex\ conditions. During \prex, we limited the vertical acceptance on the detector by turning off the high voltage on all but the two PMTs on the second row from the bottom, lowering our accidental and dead time corrections. This also reduced our sensitivity to the Levchuk correction, which will be discussed in detail in Section~\ref{subsec:CREX}. 

\begin{figure}[t]
\centerline{\includegraphics[width=\linewidth]{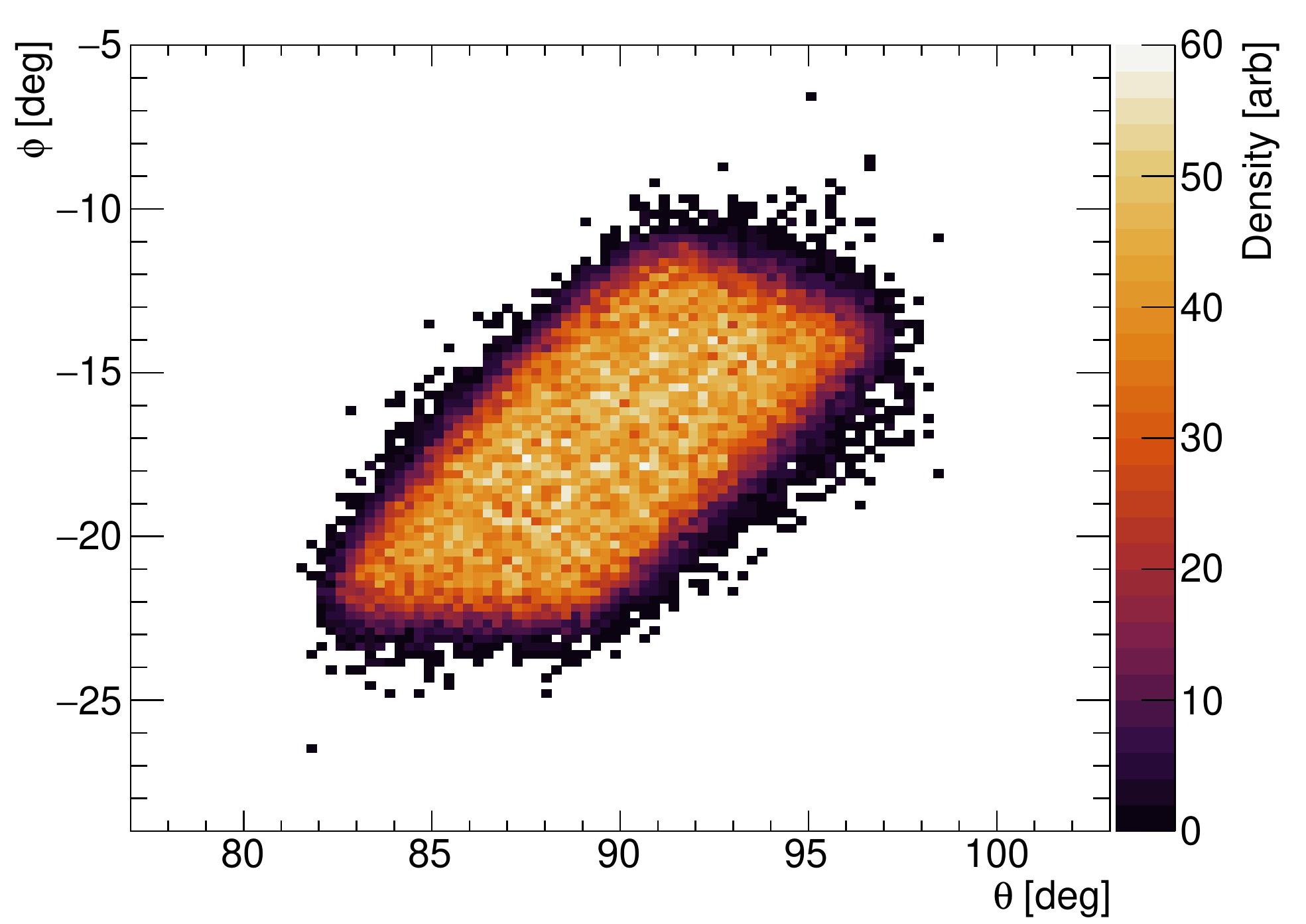}}
\caption{Heat map from our {\sc Geant4} simulataion, showing the distribution in center of mass angles during the \prex\ run, for coincident \moller\ events. Only electrons detected by the block on the left arm, second up from the bottom, in Fig.~\ref{fig:StripePREX} are included. Only the second row was active when we took data, and choosing only the left arm avoids double counting and shows the full range of $\theta$ and $\phi$.
\label{fig:CoincAnglesPREX}}
\end{figure}

Making the selection of the detector blocks on the second row leads to the accepted coincidence distribution in (center of mass angles) $\theta$ and $\phi$ shown in Fig.~\ref{fig:CoincAnglesPREX}.
Events were generated for $60^\circ\leq\theta\leq120^\circ$ and $-25^\circ\leq\phi\leq5^\circ$, where $\phi=0$ is the horizontal plane. The distribution is tilted because the holding field of the target solenoid rotates in $\phi$ by $\approx17^\circ$ at this beam energy.

We took \moller\ polarimetry runs regularly from August 4 through September 8. For the first few weeks we took data on our 4~$\mu$m thick Fe foil target until we found that it systematically produced polarizations that were 1.1\% lower than our 10~$\mu$m thick target. We found evidence that this was due to wrinkles in the thinner foil. (See Section~\ref{subsec:FoilPolarization}.) After finding this we took data on both targets for each measurement in order to systematically quantify the reliability of scaling the early polarimetry data.

\begin{figure*}
\centering
\includegraphics[width=\textwidth]{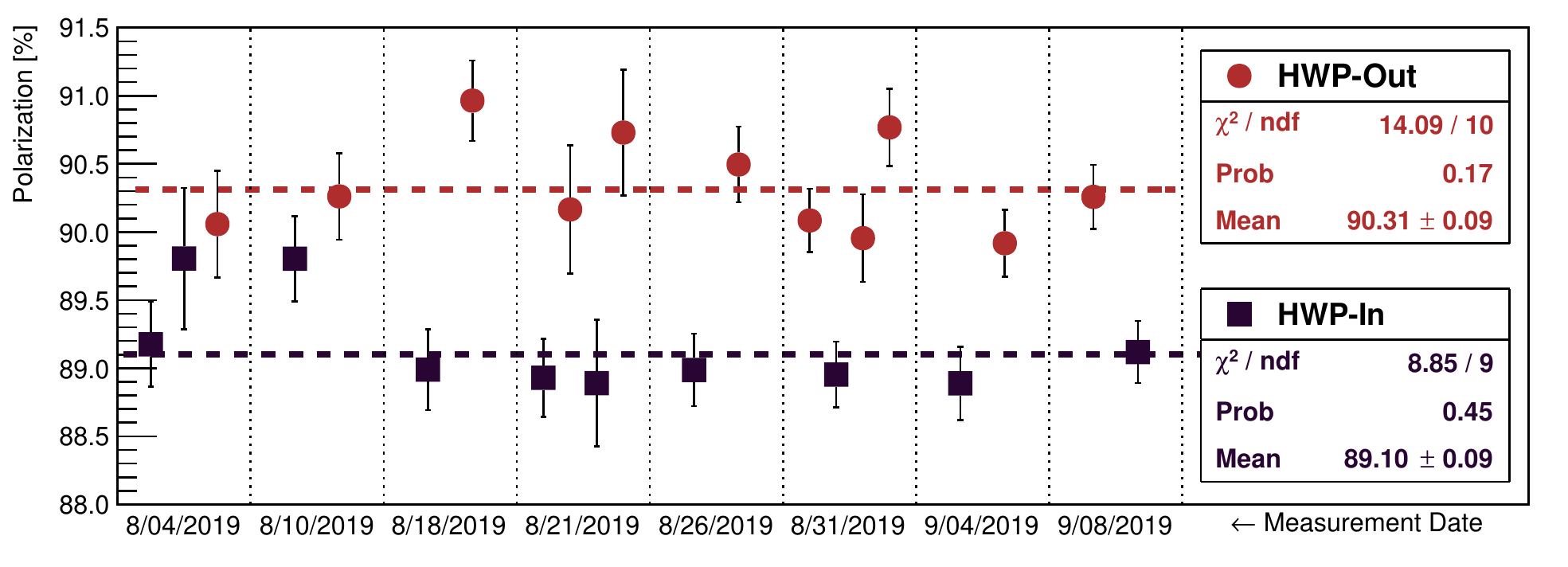}
\caption{\label{fig:prex_results}\moller\ polarization measurements taken during \prex\  separated by HWP state. The first few measurements up to and including Aug. 21 are on the 4~$\mu$m foil scaled up by 1.011 to correct for the fact that they are systematically lower than the 10~$\mu$m foil. The error bars are purely statistical.}
\end{figure*}

The full results of the polarization data for \prex\ are given in Figure~\ref{fig:prex_results}.
During parity violation experiments like \prex\ certain helicity-correlated false asymmetries are canceled by combining results where the polarization of the laser in the electron beam source is flipped by inserting a half-wave plate (HWP). During \prex, the setup produced different degrees of circular polarization when the half-wave plate was in and out. This unusual circumstance required that we report two polarizations, one for HWP in and another for HWP out. The source of this phenomenon was identified and corrected prior to running \crex.

\subsection{\moller\ polarimetry during \crex}
\label{subsec:CREX}

For \crex\ running, we built on several lessons learned from the \prex\ \moller\ polarimetry runs. These included using only the $10~\mu$m iron foil target, and correcting the laser setup at the polarized electron source so that HWP-IN and HWP-OUT should give identical beam polarizations.

\begin{figure*}
\includegraphics[width=\textwidth]{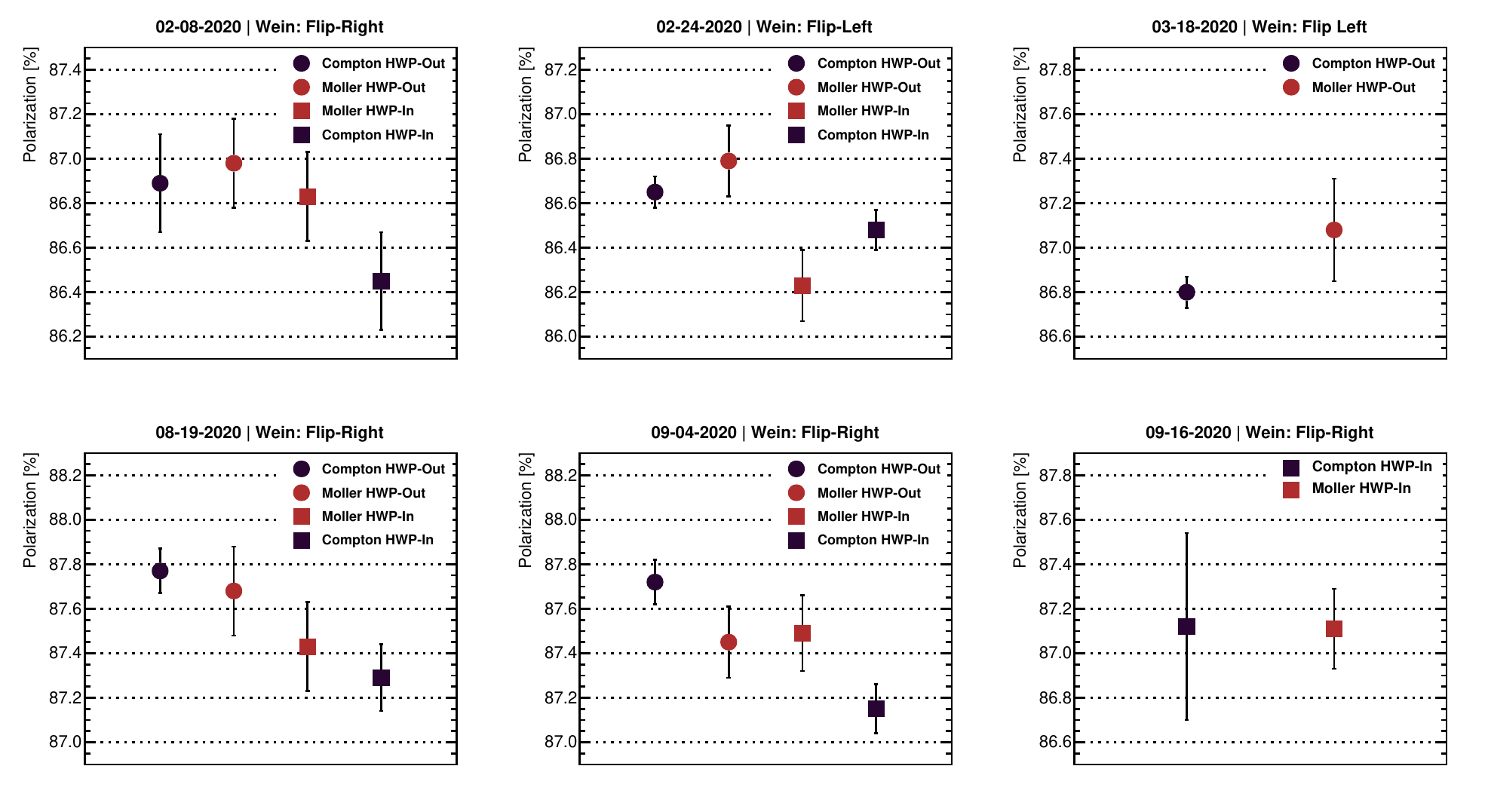}
\caption{\moller\ polarization measurements taken during \crex\ separated by the different runs, and compared to Compton polarimetry data taken just before and/or after. Error bars are purely statistical.
\label{fig:CREXresults}}
\end{figure*}

Figure~\ref{fig:CREXresults}
shows the results of our \moller\ polarimetry measurements during \crex. First notice that now there is good agreement between between the HWP-IN and HWP-OUT measurements. There is also good agreement between the flipped states of the Wien filter at the polarized electron source. Finally, given the availability of precision Compton polarimetry during \crex, there is also excellent precision agreement between the two techniques. The Compton polarimeter also confirms the equivalence of the HWP-IN and HWP-OUT states throughout the \crex\ run.

\subsection{Improved atomic momentum distributions}
\label{subsec:Kaplan}

To improve our modeling of the Levchuk Effect, we computed electron momentum distributions from tabulated Hartree-Fock one-electron wavefunctions (``orbitals'') for the isolated Fe atom \cite{bunge1993}.
The orbitals were expanded in Slater-type basis functions \cite{clementi1974} and analytically Fourier-transformed \cite{belki1989} to yield momentum-space wavefunctions $\phi_{\alpha\sigma}(\bm{p})$.
The index $\alpha$ collectively labels the orbital angular momentum and energy eigenvalue, and the spin index $\sigma=\uparrow,\downarrow$ indicates $\langle S_z \rangle = +\frac{1}{2},-\frac{1}{2}$, respectively.
We then define the electron momentum density for spin $\sigma$ as
\begin{equation}
  n_\sigma(\bm{p}) = \sum_\alpha f_{\alpha\sigma} |\phi_{\alpha\sigma}(\bm{p})|^2,
\end{equation}
where $f_{\alpha\sigma} \in \{ 0,1\}$ are occupation numbers such that $\sum_{\alpha,\sigma} f_{\alpha\sigma}=N$, the total number of electrons.
Assuming an iron target at its full saturation magnetization, all unpaired electrons in the target will have spins aligned with the saturizing field (a ferromagnetic response).
Thus the unpolarized (unp) electron momentum density is
\begin{equation}
  n_\mathrm{unp}(\bm{p}) = 2 n_\downarrow(\bm{p}),
\end{equation}
as each electron with $\langle S_z \rangle = -\frac{1}{2}$ is ``spin-paired,'' and thus the polarized (pol) electron density is
\begin{equation}
  n_\mathrm{pol}(\bm{p}) = n_\uparrow(\bm{p}) - n_\downarrow(\bm{p}).
\end{equation}
The sum of polarized and unpolarized momentum densities yields the total electron momentum density, $n(\bm{p}) = \sum_\sigma n_\sigma(\bm{p})$.
We also spherically average the momentum densities
\begin{equation}
  \langle n_\sigma(p) \rangle = \frac{ 1 }{4\pi} \int n_\sigma(\bm{p})  d \Omega_{\bm{p}}.
\end{equation}

Of course, this is not an exact calculation for magnetized iron.
The Hartree-Fock approximation neglects part of the electron-electron static Coulomb interaction, making larger errors for ``open-shell'' atoms like Fe (with configuration [Ar]$3d^6 4s^2$ \cite{bunge1993}).
Moreover, an atom within a metal, such as the Fe target, is unlike an isolated atom (e.g., the metallic bonds in solid iron produce a nearly uniform density in the interstice).
Nevertheless, the agreement in Figure~\ref{fig:ScansCREX} is quite good, suggesting that corrections to this approximation are small.

Our {\sc Geant4} simulation utilizes the momentum cumulative distribution functions (CDFs).
The CDF is derived from the momentum probability distribution function (PDF),
\begin{equation}
  \mathrm{PDF}_S(p) \equiv \left[4\pi \int_0^\infty  n_S(p) p^2 \, dp\right]^{-1} n_S(p),
  \label{eq:HFPDF}
\end{equation}
where $S=$unp, pol.
The CDF is then
\begin{equation}
  \mathrm{CDF}_S(p) \equiv 4\pi \int_0^p \mathrm{PDF}_S(p') p'^2\, dp',
  \label{eq:HFCDF}
\end{equation}
such that $\lim_{p\to \infty}\mathrm{CDF}_S(p) = 1$.

\begin{figure*}
\centering
%Side-by-side figure or top-over-bottom doesn't matter.
\includegraphics[width=0.49\textwidth]{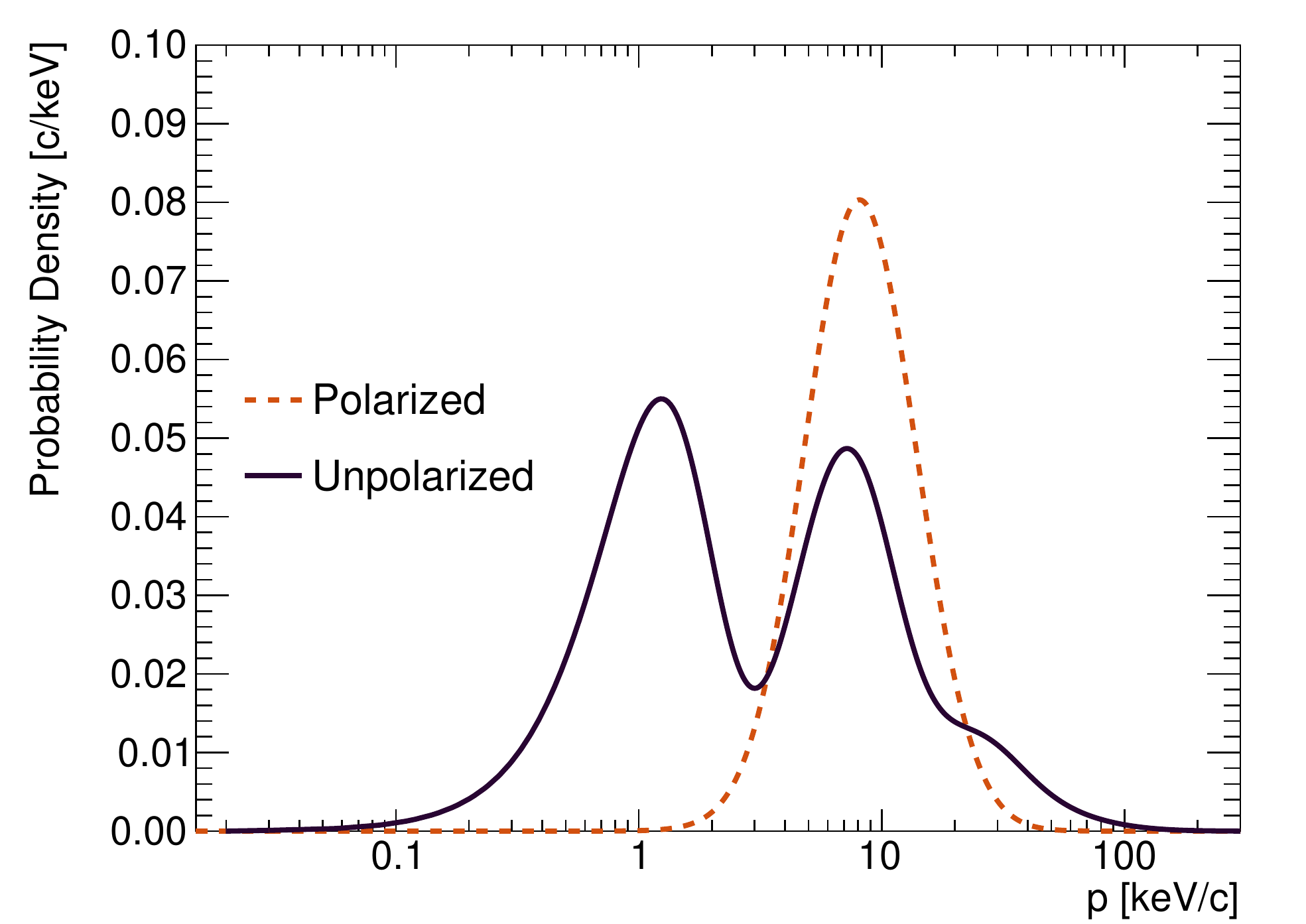}
\includegraphics[width=0.49\textwidth]{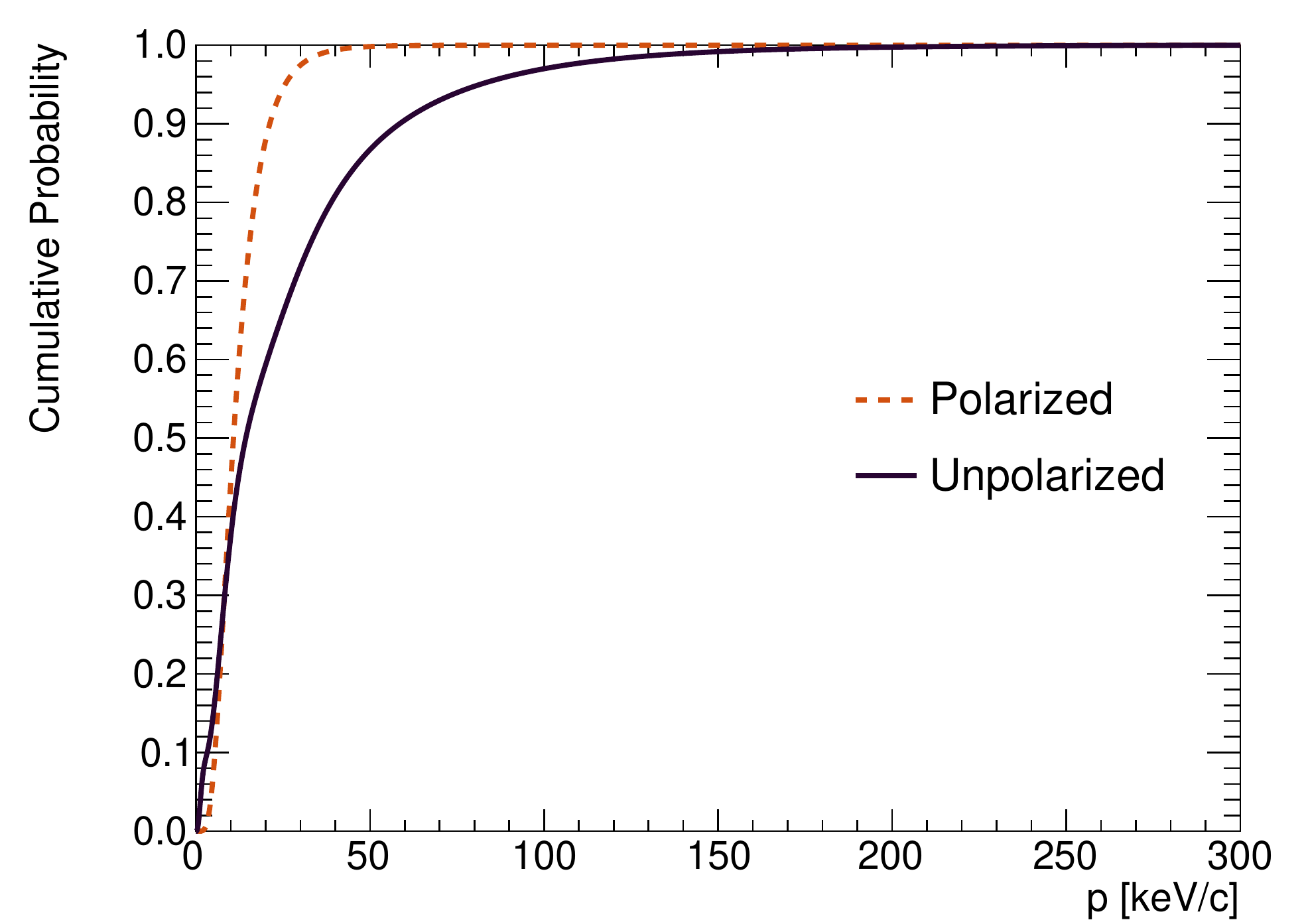}
\caption{Normalized Hartree-Fock momentum probability distribution functions $\mathrm{PDF}_S(p)$ (Eq.~\ref{eq:HFPDF}) and cumulative probability distribution functions $\mathrm{CDF}_S(p)$ (Eq.~\ref{eq:HFCDF}), separated into unpolarized and polarized components. The PDF would be compared to the ``modified hydrogen'' distributions shown in Figure~9 in~\cite{Swartz1995} .
\label{fig:HFPDFCDF}}
\end{figure*}

\begin{figure*}
\centering
    \includegraphics[width=0.49\textwidth]{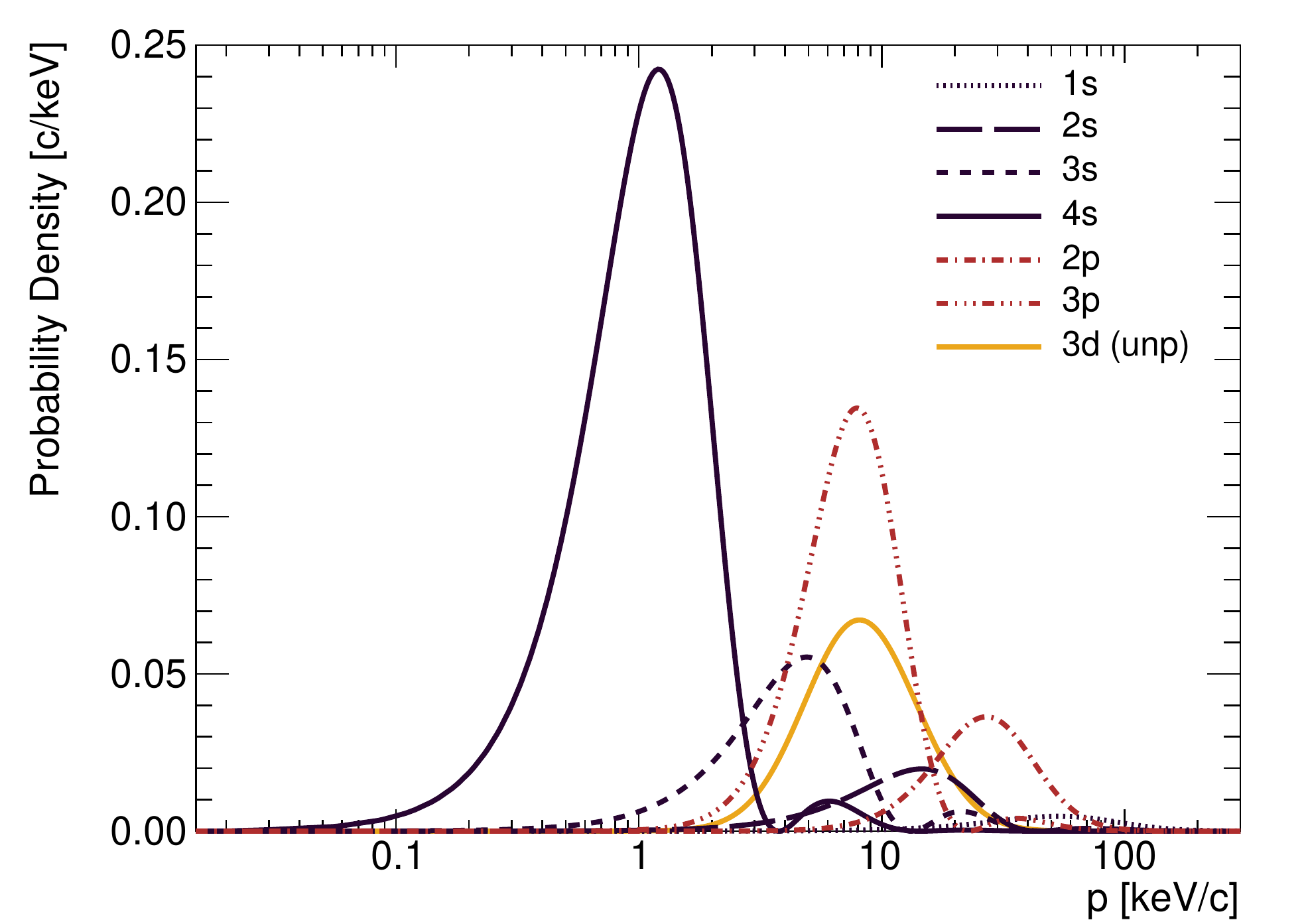}
    \includegraphics[width=0.49\textwidth]{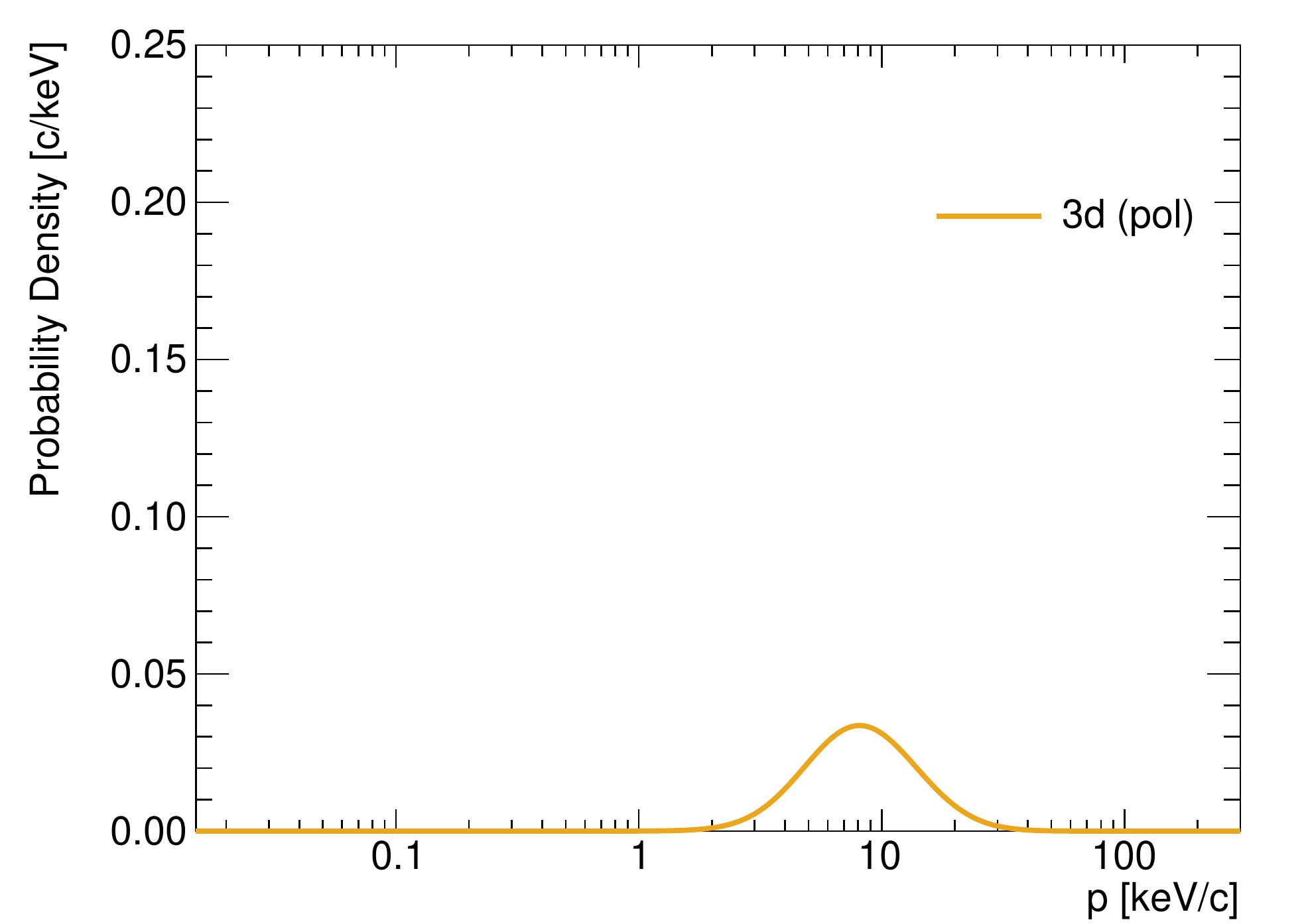}
    \caption{Contributions to the unpolarized (left) and polarized (right) momentum PDFs plotted in Fig. \ref{fig:HFPDFCDF} from each electronic orbital (shell).
    Each shell is weighted by the number of unpolarized or polarized electrons in that shell: thus a full $s$-shell would be weighted by two (zero) unpolarized (polarized) electrons, a full $p$-shell would be weighted by six unpolarized electrons, etc.
    Summing the curves in one panel would yield the corresponding momentum PDF.
    }
    \label{fig:pdf_shell_decomp}
\end{figure*}

Figure~\ref{fig:HFPDFCDF} plots $\mathrm{PDF}_S(p)$ and $\mathrm{CDF}_S(p)$ for the unpolarized and polarized electron distributions.
A shell-by-shell decomposition of the momentum PDF is plotted in Fig. \ref{fig:pdf_shell_decomp}.
Note that each shell is weighted by the number of unpolarized or polarized electrons, such that summing the contributions form each orbital would yield the corresponding momentum PDF.
The low-momentum peaks in the unpolarized PDF are largely due to $s$-shells, and the intermediate-momentum peaks are due to $p$-shells.
The Python code needed to produce the electronic wavefunctions in position and momentum space is made publicly available \cite{hf_code_repo}.

\subsection{Summary of systematic uncertainties}
\label{subsec:SystematicUncertainties}

The error bars shown in Figures~\ref{fig:prex_results} and~\ref{fig:CREXresults} are statistical only. However, our final quoted values for the beam polarization are dominated by systematic uncertainties. 

Table~\ref{tab:uncertainties}
\begin{table}
\caption{\label{tab:uncertainties}
Summary of the systematic uncertainties (in percent of the beam polarization) for \moller\ polarimetry during the \prex\ and \crex\ experiments. The total is given as a quadrature sum of the listed values. Details on how these uncertainties were determined are given in Sections~\ref{sec:measurement} and~\ref{sec:extrapolation}.}
\vspace*{0.25in}
\centering
\begin{tabular}{l c c}\hline
Uncertainty & \prex & \crex \\\hline
$\langle A_{zz}\rangle$ & 0.20 & 0.16 \\
Beam Trajectory & 0.30 & 0.00 \\
Foil Polarization & 0.63 & 0.57 \\
Dead Time & 0.05 & 0.15 \\
Charge Normalization & 0.00 & 0.01 \\
Leakage Currents & 0.00 & 0.18 \\
Laser Polarization & 0.10 & 0.06 \\
Accidentals & 0.02 & 0.04 \\
Current Dependence & 0.42 & 0.50 \\
Aperture Transmission & 0.10 & 0.10 \\
Null Asymmetry & 0.12 & 0.22 \\
July Extrapolation & 0.23 & -- \\
\hline
Total&0.89&0.85\\\hline
\end{tabular}
\end{table}
summarizes the levels we achieved for the various systematic uncertainties during \prex\ and \crex. Since \prex\ ran first, \crex\ was able to benefit from various improvements that we realized needed to be made.
The net result is that we reduced the systematic uncertainty below 1\% in each of the two experiments, with a slightly more precise result overall for \crex. The target polarization uncertainty dominates, coming in part from the precision with which we know the foil magnetization at saturation, as well as uncertainty in the degree of saturation. We reduced the uncertainty in the analyzing power $\langle A_{zz}\rangle$ largely from an improved simulation that included accurate momentum distributions for the target electrons.

\section{\label{sec:measurement}Measurement Uncertainties}
One class of systematic uncertainty includes those that pertain specifically to measurements germane to the interpretation of the \moller\ scattering asymmetry, as well as uncertainties in our scattering measurements as we made them. This section treats this class of systematic uncertainty.

\subsection{Foil Polarization}
\label{subsec:FoilPolarization}

The single largest contribution to the systematic error for \moller\ polarimetry for the \prex\ and \crex\ experiments was from target foil polarization. This error comes from uncertainty in the value for the saturation spin polarization for Fe in addition to effects such as foil alignment and flatness that impede full saturation.

Indeed, the tacit assumption of ``perfect'' knowledge of the magnetization of saturated iron was the basis for this approach to \moller\ polarimetry.~\cite{bever:1997kvc}. In a previous paper, we have carried out a study of the literature to establish this value of magnetization and quantify the uncertainties~\cite{Jones:2022jhy}, an analysis which was completed in the time since the first \prex\ publication.

Combining results of published measurements of magnetization and the gyromagnetic ratio for iron allowed us to calculate the saturation spin polarization which we found to be 8.014$\pm$0.022\% at room temperature (294 K). With beam on the target, the foil heats up a few degrees, slightly reducing the polarization. For \prex\ and \crex\ we typically ran with a 0.6~$\mu$A beam which we calculated to produce an 11$\pm3^\circ$C temperature rise, reducing the target polarization to 8.005$\pm$0.022\%. Using the modified values from~\cite{Jones:2022jhy} yields a slightly larger target polarization at room temperature of 8.020$\pm$0.018\% but with a slightly reduced calculated target heating of only 7$^\circ$C at 0.6~$\mu$A. These give a beam-on target polarization of 8.014$\pm$0.018\%, a difference of only 0.009 percentage points from the value used for \prex\ and \crex\ which is well within the quoted uncertainty.

Although the uncertainty from saturation spin target polarization is relatively small, this is, in fact, a lower bound for the actual polarization since we must also determine how close we are to reaching magnetic saturation in the foils. Semiclassical calculations by E. Stoner (see Eq. 4.27 of \cite{Stoner1950}) suggest that the alignment of the foil relative to the field is key to reaching 100\% saturation.

\begin{figure}[t]
\centering
\includegraphics[width=\linewidth]{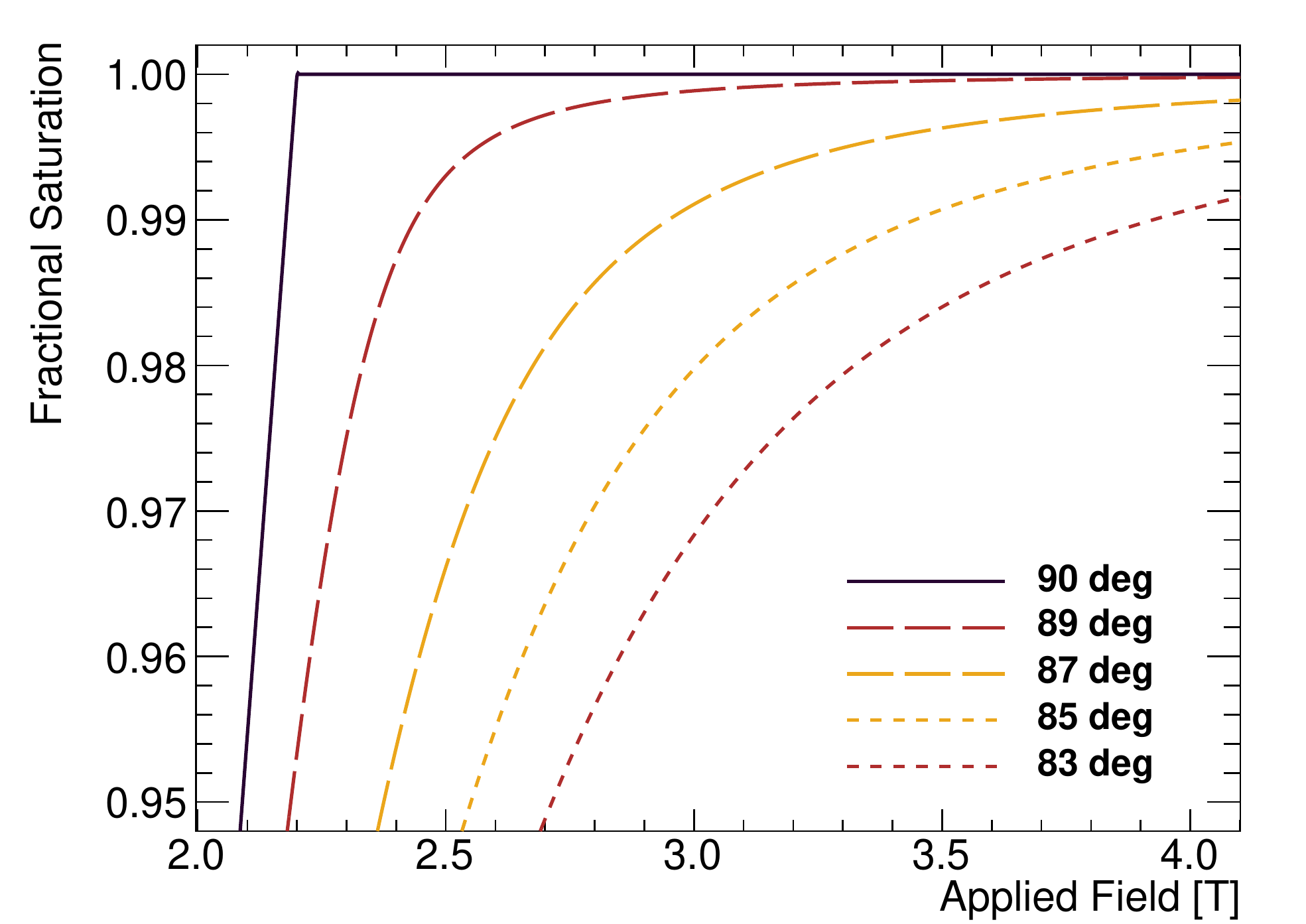}
\caption{\label{fig:stoner_curves}Semiclassical calculations~\cite{Stoner1950,Stoner:1948zz} of the fractional magnetization relative to saturation magnetization versus applied magnetic field for three different foil angles relative to the magnetic field. 90$^\circ$ is with the foil aligned perfectly normal to the field.}
\end{figure}

Figure \ref{fig:stoner_curves}
gives the results of Stoner's calculation showing for various foil angle orientations relative to the magnetic field. These curves are calculated for a prolate spheroid (our foils with a diameter of 1.27~cm and a thickness of 0.0001~cm are approximated as extremely prolate spheroids). Since this model neglects among other things, crystalline defects, material strains/stresses and impurities all of which can impede saturation, we choose to limit the quantitative use of these curves until they can be empirically validated. This model limitation is most obvious in the 90$^\circ$ curve, aligned exactly normal to the magnetic field which saturates perfectly at 2.2~T, the value of the saturation induction magnetization for iron. Previous experience at Jefferson Lab indicates that at least 3~T is required to ensure saturation of a well aligned foil target in the out-of-plane configuration. 

Looking at Fig. \ref{fig:stoner_curves} we see that even in the ideal case, a foil misalignment of a few degrees can have a measurable effect on target polarization even at high field. During \prex\ and \crex\ we operated the \moller\ target magnet at 4~T, far above the 2.2~T theoretically required to fully saturate an ideal pure iron foil out of plane. At 4~T a misalignment of 5$^\circ$ only reaches 99.5\% polarization and 7$^\circ$ only 99\%. While we expect the foil surface to be planar and aligned perpendicular to the holding field within $\pm2^\circ$, this has not been directly verified and two observations during \prex\ and \crex\ detailed next brought this assumption into question. \\

\subsection{Wrinkles in the foil}

The circular foils used for the \moller\ polarimeter in Hall A are 0.5 inches in diameter with a holder designed to tightly stretch the foils. However, the tautness of the foils is not constant from foil to foil and occasionally there are even small wrinkles that develop during the installation process. During \prex\ we utilized two foils, one 4~$\mu$m thick and another 10~$\mu$m thick. We began for the first few weeks taking data only on the 4~$\mu$m foil. When we compared with results on the 10~$\mu$m foil we found that the 4~$\mu$m foil consistently produced asymmetries that were 1.1\% smaller.

Upon further investigation we found that the 4~$\mu$m foil had a small wrinkle running across its face. When we moved 2~mm above and below center on the foil and took measurements, the measured asymmetries were 1.08$\pm$0.25\% larger than the center measurements on average. The average asymmetry on the 10~$\mu$m foil measured over a few weeks during \prex\ was a factor of 1.0110$\pm$0.0015 larger than those measured on the 4~$\mu$m foil, consistent with the 2~mm off-center measurements on the 4~$\mu$m foil.

\begin{figure}[t]
\centering
\includegraphics[width=\linewidth]{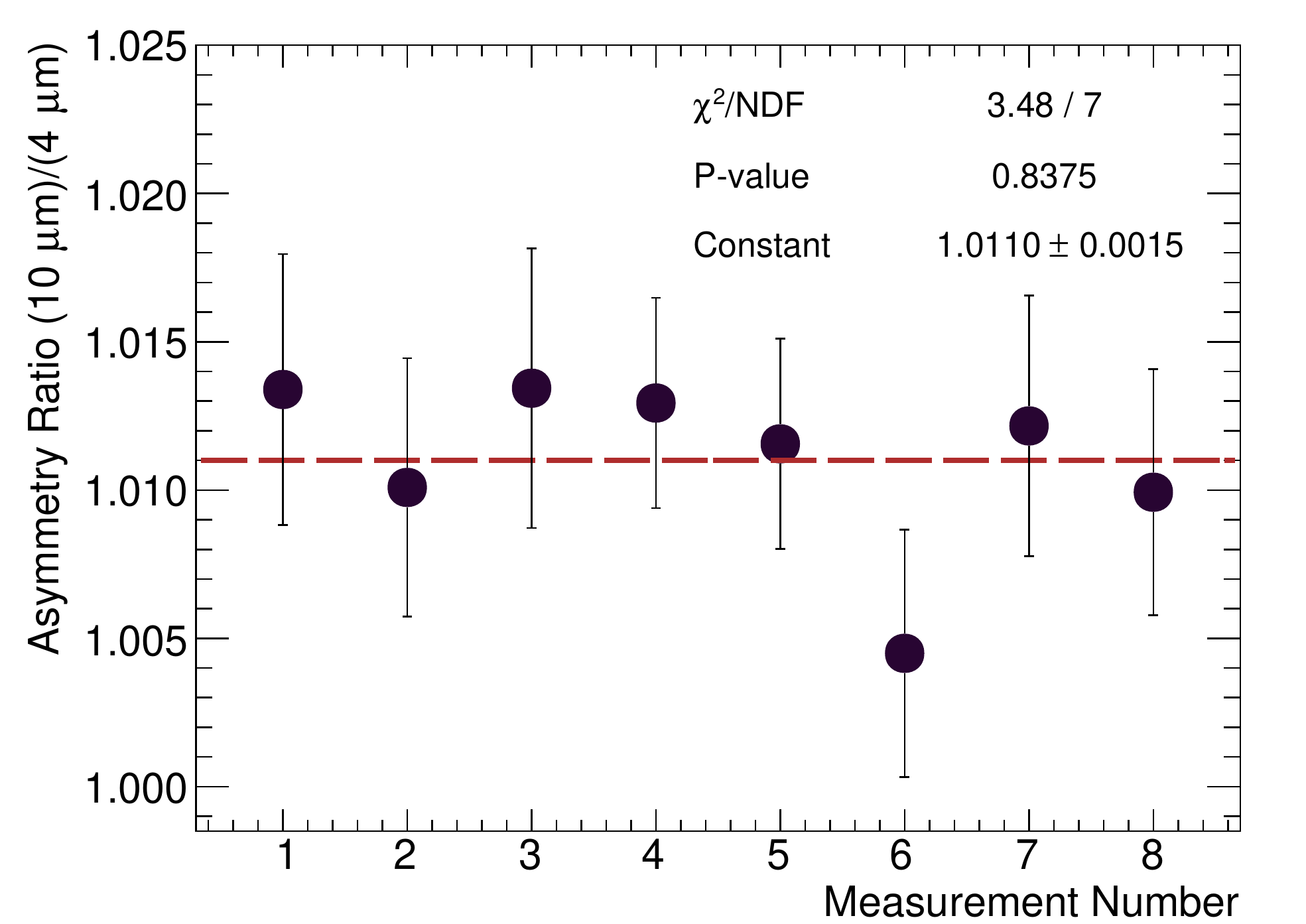}
\caption{\label{fig:10to4ratio}Ratio of measured \moller\ scattering asymmetries on the 10~$\mu$m foil to the 4~$\mu$m foil over a few weeks during \prex. We attribute the difference from unity to wrinkles in the 4~$\mu$m foil.}
\end{figure}

Figure~\ref{fig:10to4ratio} 
shows a plot of the ratios of measured asymmetries on 10 and 4~$\mu$m foils over several weeks during \prex. The average ratio, 1.011, was used to scale the earlier measurements taken during \prex\ where measurements were only taken on the 4~$\mu$m foil. For later measurements where both 10 and 4~$\mu$m data are available, the 10~$\mu$m data are used for the polarization and the 4~$\mu$m measurements are used only to find the ratio to scale the earlier data points.

The statistical consistency of the fraction over the four different measurement days included in Fig. \ref{fig:10to4ratio} provides confidence in using it to extrapolate backwards in time. However, an additional uncertainty of 0.5\% was added in quadrature with the 0.15\% statistical uncertainty in the scale factor to account for unknowns such as target and/or beam position uncertainty; however, given that only half the data used during \prex\ were taken on the 4~$\mu$m foil, the total uncertainty from this scale factor only contributes to the total polarization relative uncertainty at the $\sqrt{0.15^2+0.5^2}/2=0.26$\% level.

\subsection{Foil angle misalignment}

Although we expect the target foil ladder to be aligned normal to the magnetic holding field to within $\pm2^\circ$ or better, there is an unknown amount of warping of the foil from magnetic forces. Although theoretically, a flat foil placed precisely at the center of the magnet should have no net force on it, this is an unstable equilibrium which in practice generates rather strong torques tending to rotate the foil along the field direction. These torques are sufficiently strong that the rotation motor on the original design of the target motion system was not able to hold the target ladder perpendicular to the holding field, requiring the addition of a 50:1 gearbox for the rotation mechanism.

\begin{figure}[t]
\centering
\includegraphics[width=\linewidth]{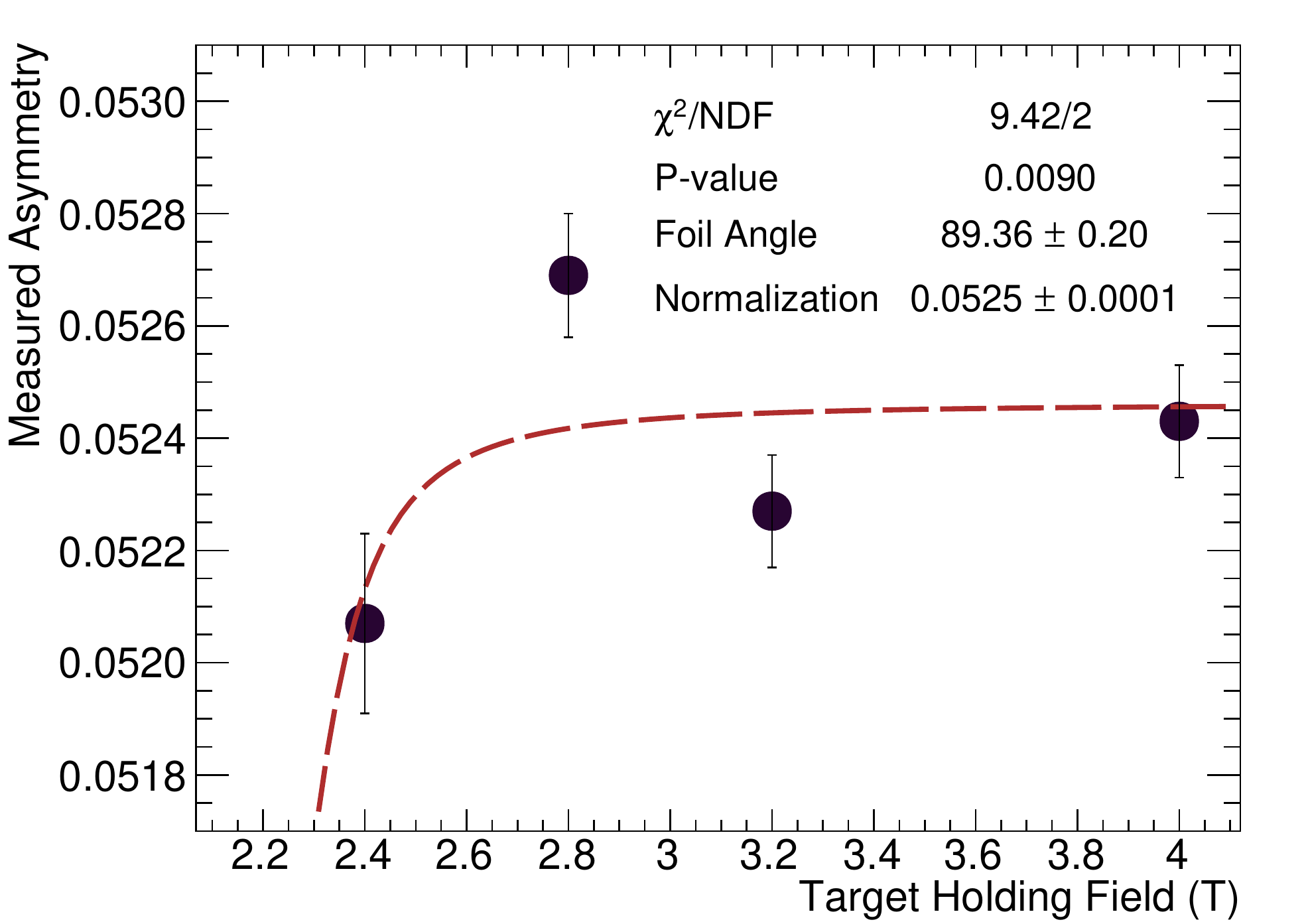}
\caption{\label{fig:measured_sat}Measured \moller\ asymmetry versus target holding field for data taken during \crex. The curve shown is the best fit to the Stoner parameterization with the angle and normalization allowed to float. }
\end{figure}

In the end, during \prex\ and \crex\ the rotation capability was disabled and the ladder bolted at a fixed orientation to ensure it remained fixed. However, given the strength of these torques, even the surface normal of a taut foil could conceivably rotate by warping or stretching with the largest effect likely at the foil center where the electron beam interacts. Although further studies are required to quantify the degree of rotation, a series of measurements taken during \crex\ did not provide satisfactory proof of full saturation even at 4~T. These measurements, seen in Fig.~\ref{fig:measured_sat},
were of the \moller\ asymmetry as a function of target holding field. For a saturated foil, one would expect an asymptotic approach to a constant value at high fields. Instead, we see that the implied polarization at 2.8~T is significantly higher than that measured at 3.4~T and 4~T. Fitting the Stoner parameterization to the data points yields a foil angle of 89.4$^\circ$ but the p-value is only 0.9\%. These data are difficult to interpret and more investigation is required to ensure saturation is achieved for future precision experiments. For both \prex\ and \crex\ an uncertainty of $\pm$0.5\% was assigned to account for uncertainty related to degree of foil saturation. 

 A total relative uncertainty of $\pm$0.63\% was assigned to foil polarization for \prex\ including 0.28\% for target saturation polarization, 0.5\% for incomplete saturation, and 0.26\% for scaling the 4~$\mu$m foil data to account for running on a warped region of the foil. For \crex\ the same uncertainties were assigned with the exception of the wrinkle uncertainty since all measurements were taken on the 10~$\mu$m foil giving a total uncertainty from foil polarization of $\pm$0.57\%.
 
 \subsection{Data Acquisition Dead Time}
 
The \moller\ polarimeter uses a standard pulse-counting data acquisition (DAQ) system, recording single and coincident discriminated signals from the electron detectors. Dead time is inherent in such systems, since another signal cannot be accepted for some period of time after the trigger. The resulting dead time correction leads to a first order correction to the asymmetry, namely 
\begin{equation}
%A_{\rm raw}=A_{\rm true}(1-{\cal R}\tau),
A_{\rm true}=A_{\rm raw} + {\cal R}\tau A_{\cal R},
\end{equation}
where $A_{\rm raw}$ is the measured raw \moller\ asymmetry, ${\cal R}$ is the singles rate average between two helicities, $\tau$ is the constant dead time of the DAQ system, and $A_{\cal R}$ is the singles rate asymmetry between two helicities.

\begin{figure}[t]
\centering
\includegraphics[width=\linewidth]{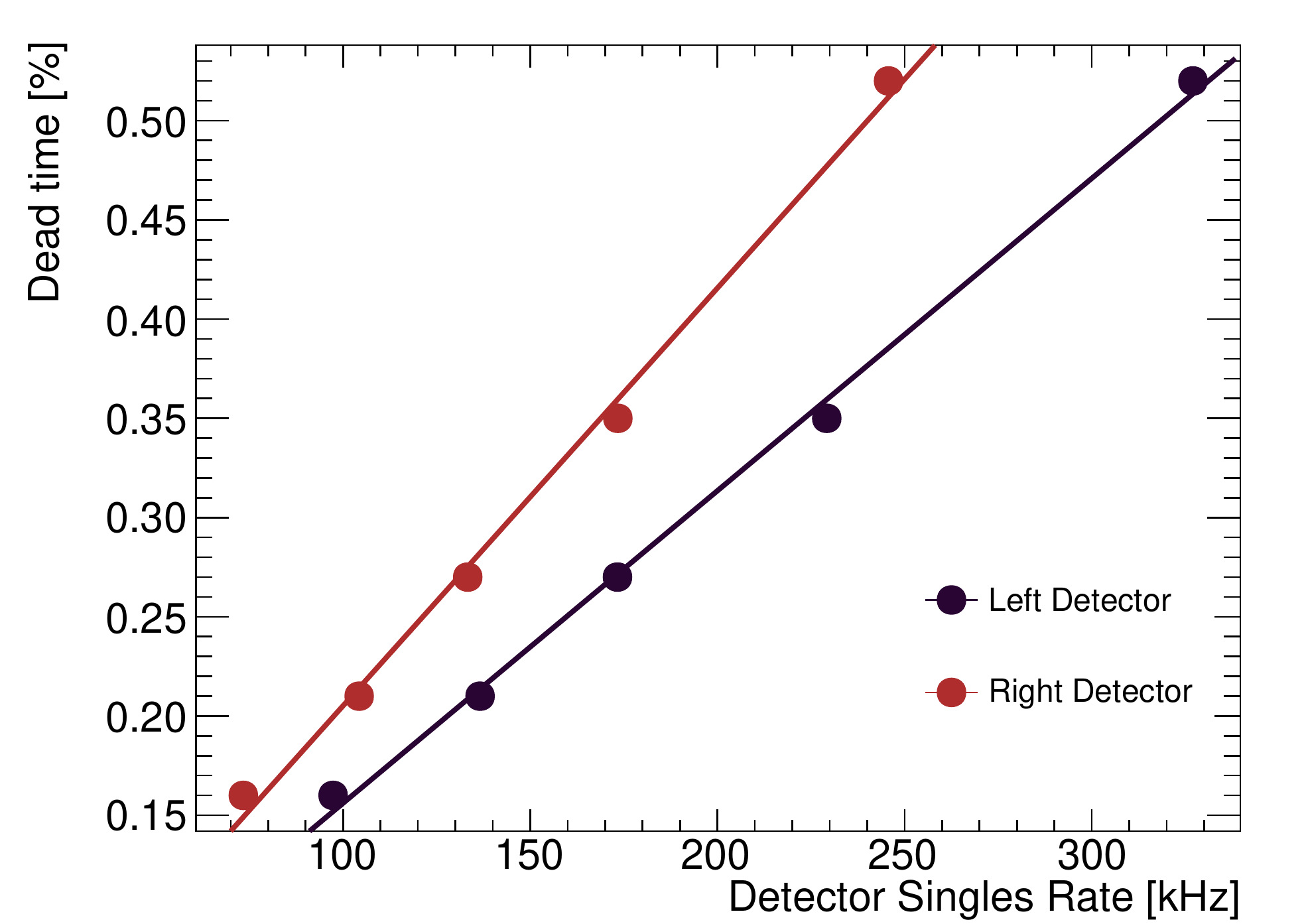}
\caption{\label{fig:deadtime}Measured fractional dead time versus detector singles rates taken using the LED pulser system during \prex.}
\end{figure}

Dead time has been measured in the past for this polarimeter with a legacy LED system that flashes at 4~kHz. The LED system is run with electron beam on the target and the dead time is calculated using the difference between the number of LED pulses and that observed via a triple coincidence between left and right detectors and the pulse generator. A measurement of dead time versus detector rate using this LED system was taken on the 10~$\mu$m foil during \prex\  with beam currents ranging from 0.4 to 1.5~$\mu$A and is shown in Fig. \ref{fig:deadtime}. This linear relationship was used to calculate the dead time as a function of detector rate for each asymmetry measurement.

For \prex\ the average correction was 0.05\% of the measured asymmetry and for \crex\ it was 0.15\%. Although this is a fairly straightforward measurement given that the legacy LED system has not yet been directly verified, we assign a 100\% uncertainty to the dead time correction.

\subsection{Accidental Correction}
When an event triggers in either the left or right detector, the coincidence logic looks for an event occurring within a few nanoseconds in the opposite detector. There is a probability that a random uncorrelated event will fall into this coincidence window. The effect of these ``accidental" coincidences on the measured asymmetry must be removed. For the Hall A \moller\ this correction is made by subtracting the coincidence rate between the right detector and a delayed ($\approx$100~ns) left detector signal. For each asymmetry the coincidence rates of both left and right helicities are decremented by the rates measured in their respective accidental coincidence windows. For \prex\ the average accidental coincidence correction was 0.12\% and for \crex\ it was 0.21\% to which we assigned a 20\% relative uncertainty. 
\subsection{Charge Normalization Correction}
To remove the effect of helicity-correlated beam current, the scattering rate for each helicity state is normalized to the simultaneously measured current. Although the beam current monitor (BCM) used to measure the current is expected to be linear over the range of the measurements, the linearity of the BCM readout used in the \moller\ DAQ was not specifically characterized during \prex\  and \crex. We estimated the size of the non-linearity by comparing the charge asymmetry as measured by the well-calibrated main experiment production DAQ system with that measured in the \moller\ DAQ system during special runs where the charge asymmetry was intentionally increased. During \prex\ the results between the two DAQ systems were consistent while measurements during \crex\ indicated that the \moller\ DAQ measured a 32\% higher asymmetry. Therefore, no uncertainty was assigned for charge normalization during \prex. However, during \crex, the charge asymmetry normalization correction to the un-normalized asymmetry was 0.029\% with a small uncertainty of 0.01\%.

\subsection{Null Asymmetry Correction}
``Null asymmetry" measurements were taken on a thin, pure Cu foil along with each \moller\ measurement on polarized Fe. Since Cu is weakly diamagnetic, to the accuracy we are concerned about, the Cu foil can be considered to have zero spin polarization. These null measurements are intended to monitor for false asymmetries arising from helicity-correlated beam properties not related to polarization (like position, angle and energy) as well as effects such as differential clipping of the beam on apertures. The average null asymmetry measured during \prex\ was $0.035\pm0.087\%$ of the measured asymmetry, consistent with zero and yielding a $1\sigma$ uncertainty of 0.12\%. During \crex\ the null asymmetry was also consistent with zero at $0.12\pm0.10\%$ of the measured asymmetry yielding a $1\sigma$ bound on a non-zero null asymmetry of 0.22\%.

\section{\label{sec:extrapolation}Extrapolation Uncertainties}
A number of potential errors arise in determining the beam polarization using \moller\ polarimetry due to the fact that these measurements are not taken at the same time and conditions as the experiment. We term these extrapolation errors and they include the extrapolation from the low current at which \moller\ measurements are taken to the high current of experimental running and a number of ``snapshot" uncertainties arising from the fact that the polarization can change in between the rather infrequent \moller\ measurements.

We note that ``snapshot uncertainties" arise because the \moller\ measurements represent only a snapshot of the beam polarization and not necessarily the time average. Even if the \moller\ polarization measurements are extremely accurate, these additional sources of error have the potential to be significant contributions when extrapolating to provide an average polarization for an experiment. To limit the uncertainty of snapshot errors, experimental collaborations can increase the frequency of measurements. Usually the frequency for \moller\ measurements is about once per week, but this can be increased if there is evidence that the polarization is changing with time. Another method sometimes employed is to use the \moller\ polarimeter results to normalize the continuous and non-invasive measurements from a Compton polarimeter. In this case the time-dependence is tracked with the Compton and the precise results of the \moller\ only are used to calibrate the overall magnitude, removing the effect of this class of uncertainty.  

%We will discuss three sources of ``snapshot errors": changes in electron source polarization from photocathode degradation, changes in electron beam precession over time, and deliberate shifts in the source laser polarization.

\subsection{High Current Extrapolation}

The largest extrapolation error for \prex\ and \crex\ was the high current extrapolation error. The strained superlattice GaAs photocathode utilized at Jefferson Lab produces the electron beam using a laser to eject electrons from the surface via the photoelectric effect. GaAs crystals alone cannot produce more than 50\% polarized electrons, but by growing layers of GaAs over layers doped with phosphorus, a lattice mismatch creates a crystal strain that removes the degeneracy of the $P_{3/2}$ and $P_{1/2}$ electrons. This small shift in relative energy removes the contribution of $P_{1/2}$ electrons, increasing the polarization so that 85-90\% polarized beams are the norm at Jefferson Lab. For a more detailed description of how these photocathodes are used to produce polarized electrons, see chapter 8 of \cite{Rao2014}. 

Further evidence of the sensitivity of the polarization to crystal strains comes from studies of exposure to atomic hydrogen\cite{Baylac2005}. They found that with sufficient exposure to hydrogen, the electron polarization could be reduced by as much as 10\%. This same overexposure of the photocathode to atomic hydrogen also was found to drastically reduce the quantum efficiency (QE) of the crystal. The authors suggest that this is due to diffusion of hydrogen into the material along the lattice mismatch region, producing an overall reduction in the crystal strain. 

Given this polarization sensitivity to photocathode crystal strains, one model for current dependence of polarization is that it could arise from injector photocathode heating. The intensity of the laser is proportional to the electron beam current, meaning that the photocathode temperature is likely to increase with beam current. The idea is that a change in temperature might change the target polarization unpredictably, for example, by changing the electronic band structure and generating phonon modes. This model can be tested by keeping the laser power on the photocathode high while limiting the beam current into the hall and comparing this result with a measurement at low laser power. 

A dedicated study in 2007 \cite{Poelker2007} used a variety of techniques to limit current to the experimental halls while increasing the laser power on the photocathode to typical experimental levels producing several tens of microamperes. We used the results of this study to limit the current dependence of the polarization to $\pm$0.42\% for \prex\ and $\pm$0.50\% for \crex. Future precision experiments such as MOLLER~\cite{MOLLER:2014iki} and SoLID~\cite{Souder:2016xcn} will require a specially designed study to further limit this dependence to the 0.1\% level. 

\subsection{Leakage Contributions \label{sec:leakage}}

Another systematic error arising from the low currents at which \moller\ measurements are taken is from current leakage from other experimental halls. The polarized electron source at Jefferson Lab was originally designed to deliver four continuous but distinct electron beams to four experimental halls or ``end stations'' A, B, C and D. This is accomplished using four co-linear pulsed lasers mode-locked at different phases to the 1497~MHz resonant frequency of the accelerator pulsing at either 499~MHz or 249.5~MHz. These electron bunches are directed by a radio-frequency (RF) field through 3 separate circular holes on an aperture plate in the injector. Although the injector lasers are pulsed, a residual baseline level of light persists between the pulses. This constant background produces a residual electric current such that the light from one hall's laser leaks into the apertures of other halls.

Given its source, the quality and polarization of this background is not expected to be the same as that of the main beam pulse. This is not usually a problem for experiments given that typical backgrounds from this source are measured in fractions of nanoamperes. However, these leakage currents can become a problem during \moller\ measurements where the current is typically less than 1~$\mu$A. During \crex,  significant leakage currents in Hall A were measured only when Hall C, the other high current hall, was operating. However, given that Hall A and C lasers in the electron source have been set up with opposite polarizations, the uncertainty from leakage currents are assessed at twice the fractional rate at which they are observed. During \prex, Hall C was not taking high enough current to produce significant leakage currents; however, during \crex\ the leakage currents from from Hall C were measured to represent on average 0.09\% of the total rate during our \moller\ measurements. We, therefore, assessed a 0.18\% uncertainty for leakage currents. 

Occasionally during past \moller\ measurements when Hall C was running at high current, the rate from leakage currents has been sufficiently high to require a reduction in the size of the main aperture in the source. This action, while reducing the background from Hall C, also slightly reduces the current from the Hall A laser and thus has the potential to change the measured polarization if, for example, there is any polarization gradient on the electron pulses. The 2007 study \cite{Poelker2007} also looked at the effect of changing the size of this aperture on the polarization and saw no statistically significant effect at the level of uncertainty of their measurements. The difference between their fully open aperture measurements and with the aperture only allowing 1-6\% transmission was 0.24$\pm$0.27\%. If an aperture dependence exists, it is below the level of sensitivity measured here. During polarization measurements we had the aperture much more open in the range of 30-100\% transmission (the main experiment runs with it wide open at 100\%) and would thus expect an even smaller sensitivity than the 2007 study. We conservatively place an additional uncertainty for this aperture dependence of $\pm$0.1\%. 

\subsection{Photocathode degradation}

The polarization of the electron beam can change with time due to changes in the quantum efficiency (QE) of the photocathode in the injector. As current is drawn off the photocathode it ionizes residual gas in the vacuum which impinges on the photocathode surface degrading its QE over time. The QE decreases with the cumulative charge pulled off its surface with a typical laser spot location yielding several hundred coulombs of charge\cite{Chao2011}. The photocathode QE is measured daily during experimental running and when the QE is sufficiently low (usually well below 0.1\%) either the laser spot location is changed or the photocathode is put through a heat and reactivation cycle to restore its QE and high polarization. Experience at Jefferson Lab has repeatedly shown that the polarization degrades with the QE particularly when the QE is low i.e. 0.1\% or less. There was no evidence of polarization degradation arising from the photocathode during \prex\ since the QE was above 0.4\% throughout the entire experiment. A single polarization was assessed for the entire experiment and no uncertainty was added for polarization changes related to photocathode degradation. 

During \crex, the charge load on the photocathode required several spot laser moves and one heat and reactivation of the photocathode creating a time-dependent evolution in the beam polarization. This evolution was accurately tracked by an optimally functioning Compton polarimeter and as a result we chose to reduce the frequency of the \moller\ measurements to every 2-3 weeks and use them only to calibrate the Compton results with a scale factor. The \moller\ results were thus used to calibrate the Compton measurements, so no additional uncertainty was assessed for polarization evolution related to photocathode degradation. \\

\subsection{Beam precession}

As the beam travels from the injector to the experiment halls at Jefferson Lab it passes through one or more arcs where dipole magnets bend the trajectory of the electrons by 180 degrees. After exiting the accelerator and just before entering Hall A, the beam goes through another 37 degree arc with the same sign of curvature. During these bends, the electrons precess predictably by an amount that depends on the net bend angle $\theta$ of the arc and the beam energy~\cite{Higinbotham2009}. The launch angle is then adjusted in the injector region such that the desired polarization (typically longitudinal) is realized in the hall. At the energy of \prex\ (950~MeV), the total precession between injector and Hall A is $\approx$300 degrees and for \crex\ (2180 MeV) it is $\approx$650 degrees (assuming an injector energy of 120~MeV and equal energy in each of the two linacs). Fluctuations in beam energy are kept well below the 0.1\% level. The uncertainty in beam polarization angle due to changes in precession resulting from fluctuations in beam energy is a fraction of a degree for both experiments and remains negligible.\\

\subsection{Changes in laser polarization}

Care is taken in setting up the source laser for PV experiments to produce close to 100\% circularly polarized light at the photocathode and to minimize differences between helicity states of the laser. The laser polarization is then actively fed back on with small shifts in the amount of linearly polarized light used to minimize residual intensity differences between helicity states. This is accomplished by incremental adjustments of the voltage across the Pockels cell that is used to flip the laser helicity. The photocathode acts as an analyzer with a slightly higher efficiency for one linear state over the other. The small shifts of a few degrees of residual linear polarization (LP) change the circular polarization (CP) a negligible amount if the initial state is close to 100\% circularly polarized (recall that for fully polarized light $\rm LP^2+CP^2=1$).

During \moller\ measurements when intensity (charge) feedback is turned off, the voltage is set to its nominal value, whereas during the experiment, it is actively adjusted to zero the charge asymmetry resulting in a slightly different average laser polarization over the experiment. During \prex\, the difference in laser polarization between \moller\ measurements and the experiment led to an overall systematic uncertainty of 0.1\%. Fortunately, during \crex\ these feedback adjustments averaged to nearly zero overall leading to a worst case 0.06\% uncertainty. 

\subsection{July extrapolation}

During July 2019, when the \prex\ experiment began taking data, the \moller\ polarimeter was still being commissioned at low energy (950~MeV). We took no reliable polarimetry measurements during this period representing about 21\% of the production data for the experiment. The events which have been observed to be associated with polarization changes at Jefferson Lab include changes in the source laser spot location on the photocathode, heat and re-activation of the photocathode, low quantum efficiency of the photocathode, deliberate change of the polarization launch angle in the injector using the Wien filter, and changes in the energy.

During \prex, the energy was deliberately stabilized to $<$0.1\% to maintain a constant acceptance on the detector; thus, energy shifts are negligible contribution to polarization uncertainty. The photocathode QE was relatively high ($>$0.4\%) throughout the experiment and there was no sign of degrading polarization even at the end of the experiment at the lowest QE when the largest effect would typically be observed. Furthermore, no spot changes on the photocathode or re-activation occurred during the experiment. The only significant polarization-altering event that occurred during the July period of data-taking was a shift in the horizontal launch angle in the injector.

\begin{figure}[t]
\centerline{\includegraphics[width=\linewidth]{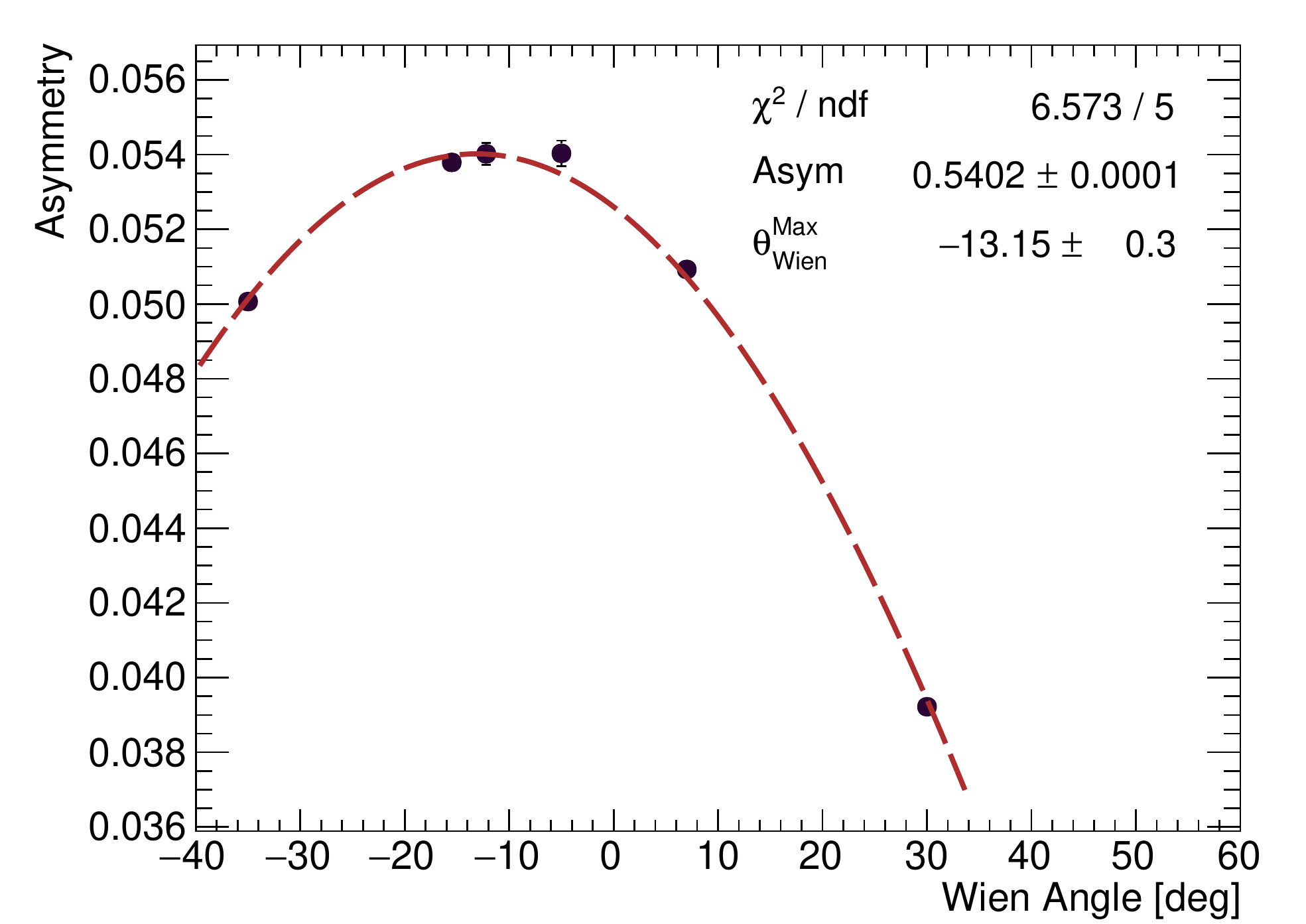}}
\caption{The result of a ``spin dance'' exercise to minimize transverse components of the beam polarization and to determine an upper bound on uncertainties arising from them. The \moller\ asymmetry is plotted as a function of the rotation angle imposed by a Wien filter at the polarized electron source. The two-parameter fit is to a cosine curve varying the amplitude and phase. The spin vector actually undergoes many $2\pi$ rotations while the electron traverses the bending magnets in CEBAF, so this technique is necessary to accurately null out transverse components when the beam is delivered to the hall.
\label{fig:SpinDance}}
\end{figure}

On July 24, 2019, a ``spin dance'' was performed where \moller\ polarization was measured as a function of injector launch angle. See Figure~\ref{fig:SpinDance}.
The largest polarization was measured to be at a horizontal Wien angle setting of 13$^\circ$, whereas prior to this we had been set at 15.5$^\circ$. In the worst case, this implies that prior to this we were running with a longitudinal polarization lower by 0.25\%. Therefore, given that  we measured no statistically significant evolution of the polarization over the remaining 79\% of the experimental data taking and we have no significant polarization-changing events in July other than the small launch angle shift, we conservatively assign an additional $\pm$1 percentage point of uncertainty to the July data and use the constant polarization determined from measurements during the remainder of the experiment.

Weighting this uncertainty with 0.21 from the amount of data without polarimetry gives a total relative uncertainty contribution from backward extrapolation to July of $\pm$0.23\%.

\section{Conclusion}
The \moller\ polarimeter in Hall~A at Jefferson Lab has been demonstrated to measure the incident electron beam polarization to better than 1\% precision, completely dominated by a number of systematic uncertainties from different sources. The largest contribution is from our knowledge of the target polarization, which includes the effects of the target angle with respect to the holding field, as well as our knowledge of the saturation magnetization of pure iron as determined from the world's available data.

The second largest contribution comes from our knowledge of the acceptance-averaged analyzing power of the spectrometer. 
%We were able to demonstrate a good understanding of the Levchuk Effect which significantly minimizes corrections that need to be made due to the motion of the inner shell electrons in iron.
We were able to demonstrate an improved understanding of the Levchuk Effect which significantly reduces systematic uncertainty considerations due to the motion of the inner shell electrons in iron.

%Our measurements show excellent agreement with the Hall~A Compton polarimeter that operated over the course of \crex.
During CREX, our measurements showed excellent agreement with the Hall~A Compton polarimeter which operated over the course of the experiment. The agreement with this second independent measurement provided persuasive evidence that our technique for restoring the spectrometer's optical tune before each measurement was effective. 

Future work will be needed to decrease these various systematic uncertainties further, in order to meet the even more stringent requirements of MOLLER~\cite{MOLLER:2014iki} and SoLID~\cite{Souder:2016xcn}. Without new measurements of the saturation magnetization of iron, we will be ultimately limited by that quantity. However, strategies to study the foil saturation curve and reduce wrinkling are being developed now. We are also working towards more direct measurements of the dead time correction and other small effects.

\section*{Acknowledgements}

The authors would like to thank the entire \prex\ and \crex\ collaborations for all of their input and a thoroughly productive endeavor, as well as the CEBAF staff. We extend a special thanks to S.\ Malace for her expertise and dedication in operating the \moller\ polarimeter throughout the experimental runs. This work was supported under US Department of Energy grants
DE-SC002042,
DE-SC0012575,
DE-FG02-07ER41522,
DE-FG02-05ER41372,
DE-AC05-06OR23177,
and
DE-FG02-84ER40146.
%DE-SC0020422 (J.N.),
%DE-SC0012575 (A.D.K.),
%DE-FG02-07ER41522 (K.D.P.),
%DE-FG02-05ER41372 (C.G.),
%and
%DE-FG02-84ER40146 (P.S.).

% \bibliography already makes a ``References'' section header, commented this out for now - AK
%\section*{References}
\bibliography{mybibfile}

\begin{thebibliography}{10}
\expandafter\ifx\csname url\endcsname\relax
  \def\url#1{\texttt{#1}}\fi
\expandafter\ifx\csname urlprefix\endcsname\relax\def\urlprefix{URL }\fi
\expandafter\ifx\csname href\endcsname\relax
  \def\href#1#2{#2} \def\path#1{#1}\fi

\bibitem{Magee:2016xqx}
J.~A. Magee, et~al., {A novel comparison of M\o{}ller and Compton electron-beam
  polarimeters}, Phys. Lett. B 766 (2017) 339--344.
\newblock \href {http://arxiv.org/abs/1610.06083} {\path{arXiv:1610.06083}},
  \href {http://dx.doi.org/10.1016/j.physletb.2017.01.026}
  {\path{doi:10.1016/j.physletb.2017.01.026}}.

\bibitem{Glamazdin:1999gg}
A.~V. Glamazdin, et~al., {Electron beam \moller\ polarimeter at JLab Hall A},
  Fizika B 8 (1999) 91--95.
\newblock \href {http://arxiv.org/abs/hep-ex/9912063}
  {\path{arXiv:hep-ex/9912063}}.

\bibitem{Steiner:1998gf}
P.~Steiner, A.~Feltham, I.~Sick, M.~Zeier, B.~Zihlmann, {A high-rate
  coincidence \moller\ polarimeter}, Nucl. Instrum. Meth. A 419 (1998)
  105--120.
\newblock \href {http://dx.doi.org/10.1016/S0168-9002(98)00938-3}
  {\path{doi:10.1016/S0168-9002(98)00938-3}}.

\bibitem{Band:1997ee}
H.~R. Band, G.~Mitchell, R.~Prepost, T.~Wright, {A \moller\ polarimeter for
  high-energy electron beams}, Nucl. Instrum. Meth. A 400 (1997) 24--33.
\newblock \href {http://dx.doi.org/10.1016/S0168-9002(97)00984-4}
  {\path{doi:10.1016/S0168-9002(97)00984-4}}.

\bibitem{Arrington:1992xu}
J.~Arrington, E.~J. Beise, B.~W. Filippone, T.~G. O'Neill, W.~R. Dodge, G.~W.
  Dodson, K.~A. Dow, J.~D. Zumbro, {A Variable energy \moller\ polarimeter at
  the MIT-Bates Linear Accelerator Center}, Nucl. Instrum. Meth. A 311 (1992)
  39--48.
\newblock \href {http://dx.doi.org/10.1016/0168-9002(92)90849-Y}
  {\path{doi:10.1016/0168-9002(92)90849-Y}}.

\bibitem{Levchuk1994}
L.~Levchuk,
  \href{https://www.sciencedirect.com/science/article/pii/0168900294905053}{The
  intra-atomic motion of bound electrons as a possible source of the systematic
  error in electron beam polarization measurements by means of a m{\o}ller
  polarimeter}, Nuclear Instruments and Methods in Physics Research Section A:
  Accelerators, Spectrometers, Detectors and Associated Equipment 345~(3)
  (1994) 496--499.
\newblock \href
  {http://dx.doi.org/https://doi.org/10.1016/0168-9002(94)90505-3}
  {\path{doi:https://doi.org/10.1016/0168-9002(94)90505-3}}.
\newline\urlprefix\url{https://www.sciencedirect.com/science/article/pii/0168900294905053}

\bibitem{bever:1997kvc}
L.~V. De~bever, J.~Jourdan, M.~Loppacher, S.~Robinson, I.~Sick, J.~Zhao, {A
  target for precise M\o{}ller polarimetry}, Nucl. Instrum. Meth. A 400~(2-3)
  (1997) 379--386.
\newblock \href {http://dx.doi.org/10.1016/S0168-9002(97)00961-3}
  {\path{doi:10.1016/S0168-9002(97)00961-3}}.

\bibitem{Jones:2022jhy}
D.~C. Jones, J.~Napolitano, P.~A. Souder, D.~E. King, W.~Henry, D.~Gaskell,
  K.~Paschke, {Accurate Determination of the Electron Spin Polarization In
  Magnetized Iron and Nickel Foils for M\o{}ller Polarimetry}, ~~\href
  {http://arxiv.org/abs/2203.11238} {\path{arXiv:2203.11238}}.

\bibitem{PREX2_2021}
D.~Adhikari, et~al., {Accurate Determination of the Neutron Skin Thickness of
  $^{208}$Pb through Parity-Violation in Electron Scattering}, Phys. Rev. Lett.
  126~(17) (2021) 172502.
\newblock \href {http://arxiv.org/abs/2102.10767} {\path{arXiv:2102.10767}},
  \href {http://dx.doi.org/10.1103/PhysRevLett.126.172502}
  {\path{doi:10.1103/PhysRevLett.126.172502}}.

\bibitem{CREX_2022}
D.~Adhikari, et~al., \href{https://arxiv.org/abs/2205.11593}{Precision
  determination of the neutral weak form factor of $^{48}${Ca}}, ~\href
  {http://arxiv.org/abs/2205.11593} {\path{arXiv:2205.11593}}, \href
  {http://dx.doi.org/10.48550/ARXIV.2205.11593}
  {\path{doi:10.48550/ARXIV.2205.11593}}.
\newline\urlprefix\url{https://arxiv.org/abs/2205.11593}

\bibitem{GEANT4:2002zbu}
S.~Agostinelli, et~al., {{\sc Geant}4--a simulation toolkit}, Nucl. Instrum.
  Meth. A 506 (2003) 250--303.
\newblock \href {http://dx.doi.org/10.1016/S0168-9002(03)01368-8}
  {\path{doi:10.1016/S0168-9002(03)01368-8}}.

\bibitem{Allison:2006ve}
J.~Allison, et~al., {{\sc Geant}4 developments and applications}, IEEE Trans.
  Nucl. Sci. 53 (2006) 270.
\newblock \href {http://dx.doi.org/10.1109/TNS.2006.869826}
  {\path{doi:10.1109/TNS.2006.869826}}.

\bibitem{Allison:2016lfl}
J.~Allison, et~al., {Recent developments in {\sc Geant}4}, Nucl. Instrum. Meth.
  A 835 (2016) 186--225.
\newblock \href {http://dx.doi.org/10.1016/j.nima.2016.06.125}
  {\path{doi:10.1016/j.nima.2016.06.125}}.

\bibitem{EricKingThesis}
D.~E. King, Utilizing parity violating electron scattering as a probe to
  measure the neutron radius of $^{208}${Pb}, Ph.D. thesis, Syracuse
  University, copyright - Database copyright ProQuest LLC (2021).

\bibitem{AllisonZecThesis}
A.~Zec, Compton polarimetry for neutral weak form factor measurements in
  $^{208}${Pb} and $^{48}${Ca}, Ph.D. thesis, University of Virginia (2022).

\bibitem{Afanasev:2006xs}
A.~Afanasev, E.~Chudakov, A.~Ilyichev, V.~Zykunov, {MERADGEN 1.0: Monte Carlo
  generator for the simulation of radiative events in polarized \moller\
  scattering}, Comput. Phys. Commun. 176 (2007) 218--231.
\newblock \href {http://arxiv.org/abs/hep-ph/0603027}
  {\path{arXiv:hep-ph/0603027}}, \href
  {http://dx.doi.org/10.1016/j.cpc.2006.10.002}
  {\path{doi:10.1016/j.cpc.2006.10.002}}.

\bibitem{Swartz1995}
M.~Swartz, H.~Band, F.~Decker, P.~Emma, M.~Fero, et~al.,
  \href{https://www.sciencedirect.com/science/article/pii/0168900295003843}{Observation
  of target electron momentum effects in single-arm \moller\ polarimetry},
  Nuclear Instruments and Methods in Physics Research Section A: Accelerators,
  Spectrometers, Detectors and Associated Equipment 363~(3) (1995) 526--537.
\newblock \href
  {http://dx.doi.org/https://doi.org/10.1016/0168-9002(95)00384-3}
  {\path{doi:https://doi.org/10.1016/0168-9002(95)00384-3}}.
\newline\urlprefix\url{https://www.sciencedirect.com/science/article/pii/0168900295003843}

\bibitem{bunge1993}
C.~Bunge, J.~Barrientos, A.~Bunge,
  \href{http://www.sciencedirect.com/science/article/pii/S0092640X8371003X}{Roothaan-hartree-fock
  ground-state atomic wave functions: Slater-type orbital expansions and
  expectation values for z = 2-54}, Atomic Data and Nuclear Data Tables 53~(1)
  (1993) 113 -- 162, the expansion coefficients and orbital exponents were
  taken from the table on page 134, the lowest-energy configuration for Fe in
  that work.
\newblock \href {http://dx.doi.org/https://doi.org/10.1006/adnd.1993.1003}
  {\path{doi:https://doi.org/10.1006/adnd.1993.1003}}.
\newline\urlprefix\url{http://www.sciencedirect.com/science/article/pii/S0092640X8371003X}

\bibitem{clementi1974}
E.~Clementi, C.~Roetti,
  \href{http://www.sciencedirect.com/science/article/pii/S0092640X74800161}{Roothaan-{Hartree}-{Fock}
  atomic wavefunctions: {Basis} functions and their coefficients for ground and
  certain excited states of neutral and ionized atoms, {$Z\leq$} 54}, Atomic
  Data and Nuclear Data Tables 14~(3) (1974) 177 -- 478.
\newblock \href
  {http://dx.doi.org/https://doi.org/10.1016/S0092-640X(74)80016-1}
  {\path{doi:https://doi.org/10.1016/S0092-640X(74)80016-1}}.
\newline\urlprefix\url{http://www.sciencedirect.com/science/article/pii/S0092640X74800161}

\bibitem{belki1989}
D.~Belki{\'{c}}, H.~S. Taylor,
  \href{https://doi.org/10.1088/0031-8949/39/2/004}{A unified formula for the
  {Fourier} transform of {Slater}-type orbitals}, Physica Scripta 39~(2) (1989)
  226--229.
\newblock \href {http://dx.doi.org/10.1088/0031-8949/39/2/004}
  {\path{doi:10.1088/0031-8949/39/2/004}}.
\newline\urlprefix\url{https://doi.org/10.1088/0031-8949/39/2/004}

\bibitem{hf_code_repo}
All code for evaluating the Hartree-Fock momentum distributions is publicly
  available: https://gitlab.com/dhamil/levchuk-dft-corrections (2022).

\bibitem{Stoner1950}
C.~S. Edmund, \href{https://doi.org/10.1088/0034-4885/13/1/304}{Ferromagnetism:
  magnetization curves}, Reports on Progress in Physics 13~(1) (1950) 83--183.
\newblock \href {http://dx.doi.org/10.1088/0034-4885/13/1/304}
  {\path{doi:10.1088/0034-4885/13/1/304}}.
\newline\urlprefix\url{https://doi.org/10.1088/0034-4885/13/1/304}

\bibitem{Stoner:1948zz}
E.~Stoner, E.~Wohlfarth, {A Mechanism of Magnetic Hysteresis in Heterogeneous
  Alloys}, {Phil.Trans. Royal Soc. London Series A} {240}~({826}) ({1948})
  {599--642}.

\bibitem{Rao2014}
T.~Rao, D.~H. Dowell, An engineering guide to photoinjectors (2014).
\newblock \href {http://arxiv.org/abs/1403.7539} {\path{arXiv:1403.7539}}.

\bibitem{Baylac2005}
M.~Baylac, et~al., {Effects of atomic hydrogen and deuterium exposure on high
  polarization GaAs photocathodes}, Phys. Rev. ST Accel. Beams 8 (2005) 123501.
\newblock \href {http://dx.doi.org/10.1103/PhysRevSTAB.8.123501}
  {\path{doi:10.1103/PhysRevSTAB.8.123501}}.

\bibitem{Poelker2007}
M.~Poelker, J.~Grames, J.~Hansknecht, R.~Kazimi, J.~Musson,
  \href{https://link.aps.org/doi/10.1103/PhysRevSTAB.10.053502}{Generation of
  electron bunches at low repetition rates using a beat-frequency technique},
  Phys. Rev. ST Accel. Beams 10 (2007) 053502.
\newblock \href {http://dx.doi.org/10.1103/PhysRevSTAB.10.053502}
  {\path{doi:10.1103/PhysRevSTAB.10.053502}}.
\newline\urlprefix\url{https://link.aps.org/doi/10.1103/PhysRevSTAB.10.053502}

\bibitem{MOLLER:2014iki}
J.~Benesch, et~al., {The MOLLER Experiment: An Ultra-Precise Measurement of the
  Weak Mixing Angle Using \moller\ Scattering}, arXiv\href
  {http://arxiv.org/abs/1411.4088} {\path{arXiv:1411.4088}}.

\bibitem{Souder:2016xcn}
P.~A. Souder, {Parity Violation in Deep Inelastic Scattering with the SoLID
  Spectrometer at JLab}, Int. J. Mod. Phys. Conf. Ser. 40 (2016) 1660077.
\newblock \href {http://dx.doi.org/10.1142/S2010194516600776}
  {\path{doi:10.1142/S2010194516600776}}.

\bibitem{Chao2011}
Y.~C. Chao, M.~Drury, C.~Hovater, A.~Hutton, G.~A. Krafft, M.~Poelker,
  C.~Reece, M.~Tiefenback,
  \href{https://doi.org/10.1088/1742-6596/299/1/012015}{{CEBAF} accelerator
  achievements}, Journal of Physics: Conference Series 299 (2011) 012015.
\newblock \href {http://dx.doi.org/10.1088/1742-6596/299/1/012015}
  {\path{doi:10.1088/1742-6596/299/1/012015}}.
\newline\urlprefix\url{https://doi.org/10.1088/1742-6596/299/1/012015}

\bibitem{Higinbotham2009}
D.~W. Higinbotham, {Electron Spin Precession at CEBAF}, AIP Conf. Proc.
  1149~(1) (2009) 751--754.
\newblock \href {http://arxiv.org/abs/0901.4484} {\path{arXiv:0901.4484}},
  \href {http://dx.doi.org/10.1063/1.3215753} {\path{doi:10.1063/1.3215753}}.

\end{thebibliography}

\end{document}